\documentclass[twocolumn]{aa}

\usepackage{graphicx}

\begin{document}

%%%%%%%%%%%%%%%%%%%%%%%%%%%%%%%%%%%%%%%%%%%%%%%%%%%%%%%%%%%%%%%%%%%%%%%%%%%%%%%
   \title{The transmission spectrum of Earth-size transiting planets}
   \titlerunning{Spectrum of Earth-size transiting planets}

   \author{D. Ehrenreich\inst{1} \and G. Tinetti\inst{2} \and A. Lecavelier des
   Etangs\inst{1} \and A. Vidal-Madjar\inst{1} \and F. Selsis\inst{3}}

   \offprints{D. Ehrenreich}

   \institute{
   Institut d'Astrophysique de Paris, CNRS (UMR~7095) -- Universit\'e Pierre \& Marie
   Curie,
   98 bis, boulevard Arago 75014 Paris, France, \email{ehrenreich@iap.fr}
   \and
   NASA Astrobiology Institute, California Institute of
   Technology, IPAC, MS 220-6, 1\,200 E. California, Pasadena,
   91125 (CA), USA
   \and
   Centre de Recherche Astronomique de Lyon, \'Ecole Normale Sup\'erieure,
   47, all\'ee d'Italie 69364 Lyon Cedex 7, France
   }

   \date{Received / Accepted}

\abstract{A variety of terrestrial planets with different
physical parameters and exotic atmospheres might plausibly
exist outside our Solar System, waiting to be detected by the
next generation of space-exploration missions. Some of these
planets might transit their parent star. We present here the
first study of atmospheric signatures of transiting Earth-size
exoplanets. We focus on a limited number of significant
examples, for which we discuss the detectability of some of the
possible molecules present in their atmospheres, such as water
(H$_2$O), carbon dioxide (CO$_2$), ozone (O$_3$) or molecular
oxygen (O$_2$). To this purpose, we developed a model to
simulate transmission spectra of Earth-size exoplanets from the
ultraviolet (UV) to the near infrared (NIR). According to our
calculations, the signatures of planetary atmospheres represent
an absorption of a few parts-per-million (ppm) in the stellar
flux. The atmospheres of a few Earth-like planets can be
detected with a 30--40~m telescope. The detection of the
extensive atmospheres of tens of small satellites of giant
exoplanets and hundreds of hypothetical ocean-planets can be
achieved with 20--30~m and 10--20~m instruments, respectively,
provided all these planets are frequent and they are
efficiently surveyed. We also found that planets around K stars
are favored, mainly because these stars are more numerous and
they are smaller compared to G or F stars. While not addressed
in this study, limitations might come from the stellar
photometric micro-variability.}

    \maketitle
%%%%%%%%%%%%%%%%%%%%%%%%%%%%%%%%%%%%%%%%%%%%%%%%%%%%%%%%%%%%%%%%%%%%%%%%%%%%%%%

\section{Introduction}
%=====================
The Earth is the only known example of a life-hosting world,
even though terrestrial exoplanets have been searched for since
the discovery of the first Earth-mass exoplanets by Wolszczan
\& Frail (1992). However, planets similar to the Earth, Venus
or Mars, in size, density or orbital parameters are still
beyond the reach of the present capabilities for planet
detection around normal stars.

Until now, mostly giant exoplanets have been discovered.
Remarkable progresses have been made recently with the
discovery of planets in the mass range of 14 to 21~Earth masses
(14 to 21~M$_\oplus$, see McArthur et al.\ 2004; Santos et al.\
2004), and most recently a $\sim$7.5~M$_\oplus$ planet orbiting
\object{GJ~876} (Rivera et al.\ 2005). We may speculate, then,
that smaller planets with sizes down to that of the Earth might
be observed in a near future. Among the 161
planets\footnote{From J.~Schneider's Extrasolar Planets
Encyclop\ae dia at
\texttt{vo.obspm.fr/exoplanetes/encyclo/encycl.html}. See also
the web page of the IAU Working Group on Extrasolar Planets at
\texttt{www.ciw.edu/boss/IAU/div3/wgesp}.} detected so far,
eight have been discovered or re-discovered as they were
transiting their parent star, producing a photometric
occultation. The last transiting planet identified is a
Saturn-mass planet orbiting \object{HD~149\,026}, a bright
$V=8.15$ G0\,{\sc iv} star (Sato et al.\ 2005). The first
discovered transiting giant exoplanet, HD~209\,458b (Henry et
al.\ 2000; Charbonneau et al.\ 2000; Mazeh et al.\ 2000), is
the object of intense investigations dedicated to
characterizing its hot atmosphere.

Probing planetary atmospheres by stellar occultations is an
effective method used for a lot of planets and their satellites
in the Solar System, from Venus to Charon (see, e.g., Elliot \&
Olkin 1996). With this technique, we can observe the thin
atmospheric ring surrounding the optically thick disk of the
planet: the limb. In the case of giant exoplanets, though, the
star is only partially occulted (1.6\% for the transiting
planet \object{HD~209\,458b}). The spectrum of the star light
transmitted and filtered by the lower and thick giant exoplanet
atmosphere consequently presents extremely weak absorption
features (from $10^{-3}$ to $10^{-4}$, see Seager \& Sasselov
2000; Hubbard et al.\ 2001; Brown 2001).

Despite the difficulties, such dim signatures were detected:
Charbonneau et al.\ (2002) measured the lower atmosphere of
\object{HD~209\,458b} as they detected a \mbox{$(2.32 \pm 0.57)
\cdot 10^{-4}$} photometric diminution in the sodium doublet
line of the parent star at 589.3~nm. However its upper
atmosphere, which extends up to several planet radii, shows
even larger signatures. Vidal-Madjar et al.\ (2003, 2004)
observed a \mbox{$15 \pm 4 \%$} absorption in the
Lyman~$\alpha$ (Ly$_\alpha$) emission line of
\object{HD~209\,458} at 121.57~nm as well as absorptions from
atomic carbon (\mbox{$7.5 \pm 3.5 \%$}) and oxygen (\mbox{$13
\pm 4.5 \%$}) in the upper atmosphere. In this work, we will
discuss the possibility to detect and to characterize the lower
atmospheres of exoplanets using signatures comparable in origin
to the one detected by Charbonneau et al.\ (2002).

The idea is to extend the use of transmission spectroscopy to
hypothetical Earth-size planets. We estimate that these
exoplanets present at least two orders of magnitude less signal
than gaseous giants, as the transit of the planet itself would
have a dimming of $\sim$10$^{-5}$ (the transit depth, $\Delta F
/ F$, where $F$ is the stellar flux, can be expressed as $(R_P
/ R_\star)^2$, with $R_P$ and $R_\star$ standing for the radii
of the planet and the star, respectively). The atmospheres of
Earth-size exoplanets should span over $\sim$100-km height
without considering potential upper atmospheres. Depending on
their transparency -- which would give an equivalent optically
thick layer of $\sim$10~km -- the expected occultations caused
by atmospheric absorptions should be $\sim$10$^{-7}$ to
$\sim$10$^{-6}$.

Earth-size planets are probably the most challenging objects to
detect with transmission spectroscopy. The orders of magnitude
given above, in fact, raise many questions: is it realistic to
seek for possible features that dim, with an instrumentation
that might or might not be available in a near future? What are
the strongest signatures we should expect? What kind of planet
could be the best candidate to look at?

We have developed a one-dimensional model of transmission at
the limb to give quantitative answers to these questions. Since
we use the stellar light to explore the planetary atmospheres,
we chose to focus on the wavelength range where the largest
number of photons is available, i.e., between 200 and
2\,000~nm. The model is described in Sect.~\ref{sec:model}. The
detectability of the selected atmospheres depends on the
signal-to-noise ratio (S/N) achievable with a space telescope
spectrograph. The constraints on idealized observations and the
method we used to calculate their S/N, are described in
Sect.~\ref{sec:S/N}. Finally, the results for the specified
cases are given and discussed in Sect.~\ref{sec:results}.

\section{Model description}
%==========================
\label{sec:model}

\subsection{Geometric description of the model}
%----------------------------------------------
\label{sec:geometry}

The general geometry of a transiting system is described by
Brown (2001). In the present work we consider a non-transient
occultation for the `in transit' phase, with a null phase
configuration (configuration~2 in Brown's Fig.~1), that is, the
planet is centered in the line of sight with respect to the
star. This configuration both maximizes the area of the
atmosphere that is filtering the stellar light and minimizes
any effects linked to the stellar limb darkening (Seager \&
Sasselov 2000).

\begin{figure}
\resizebox{\hsize}{!}{\includegraphics{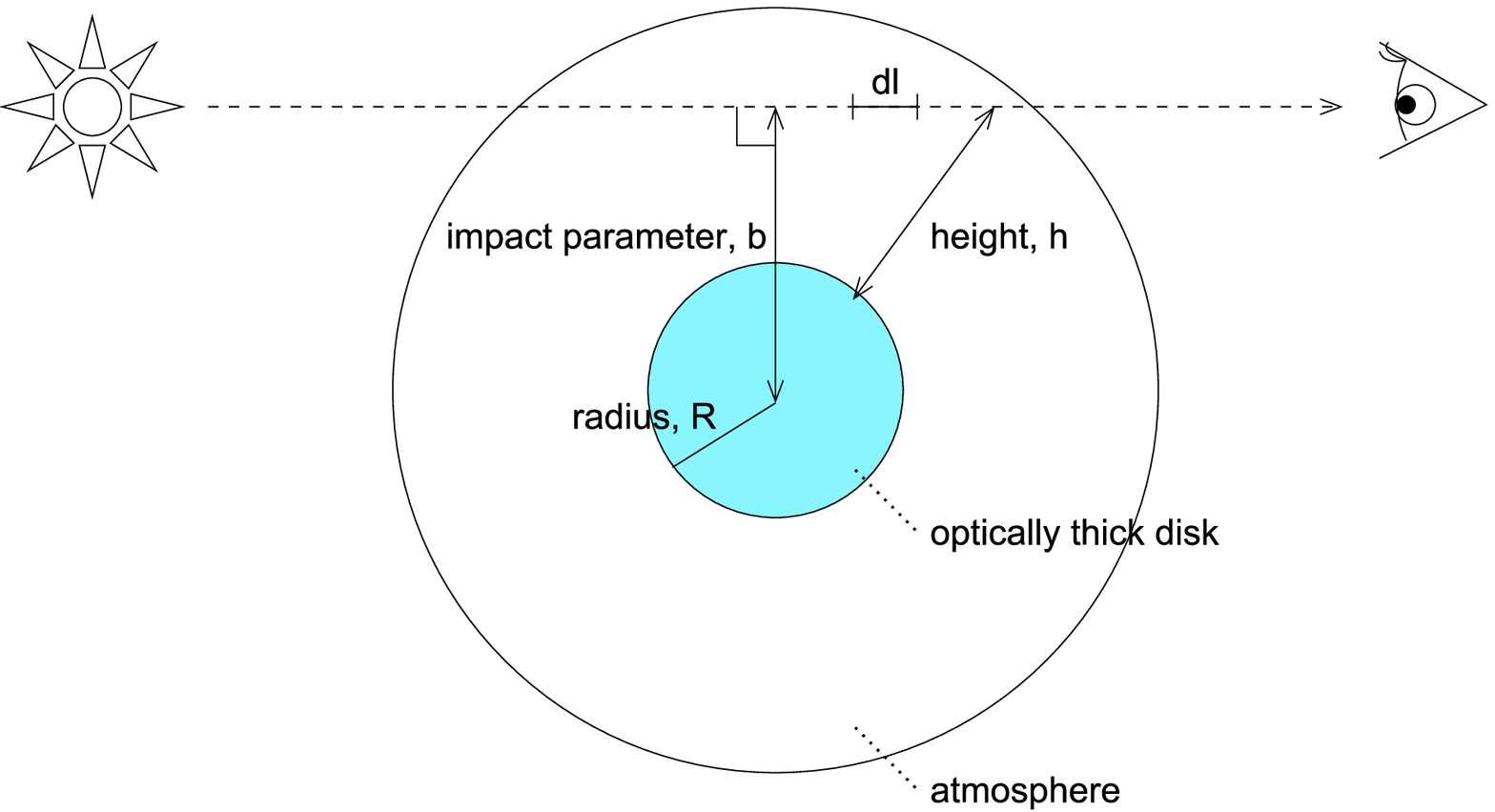}}
\caption{Sketch of the transmission of the stellar light
through the planetary limb. The planet itself, i.e. the `solid'
disk (in grey) is optically thick at all wavelengths. The
quantity $\mathrm{d}l$ is the elemental length along the line
of sight. In the calculation, we prefer to use the height $h$
instead of $l$. The scale in the figure has been distorted for
clarity.} \label{fig:transmission_geometry}
\end{figure}

The stellar light is filtered through the atmospheric limb of
the planet, as sketched in
Fig.~\ref{fig:transmission_geometry}. In the following we
detail the integration of the atmospheric opacity along a
stellar light path (or cord) through the limb of the planet.

\subsubsection{Opacity along the line of sight}
%''''''''''''''''''''''''''''''''''''''''''''''
We calculate the total opacity of the model atmosphere,
$\tau_{\lambda}$, along a cord, parallel to the line of sight,
as the sum of the opacity of each species $i$ present in the
atmosphere, $\tau_{\lambda} = \sum_{i}\tau_{\lambda,i}$. We can
calculate the opacity along the cord as a function of its
impact parameter, $b$:
\begin{equation} \label{eq:opacity}
\tau_{\lambda,i}(b) = 2 \int_{0}^{+\infty} A_{\lambda,i}
\rho_i(h) \mathrm{d} l,
\end{equation}
where $A_{\lambda,i}$ is the absorption coefficient for the
species $i$ at the wavelength $\lambda$, expressed in
cm$^2$\,g$^{-1}$, and $\rho_i(h)$ is the mass density in
g\,cm$^{-3}$ of the species $i$ at an altitude $h$ in the
atmosphere.

Now, re-expressing Eq.~\ref{eq:opacity} as a function of the
height $z = h + R_P$, with $R_P$ being the planet radius, we
obtain:
\begin{equation}
\tau_{\lambda,i}(b) = 2 \int_{b}^{b_\mathrm{max}} A_{\lambda,i}
\rho_i(z-R_P)
    \frac{z \mathrm{d} z}{\sqrt{z^2-b^2}},
\end{equation}
where $b_\mathrm{max}$ is the height of the higher atmospheric
level we are considering. The method to estimate
$b_\mathrm{max}$ is presented in Sect.~\ref{sec:b_max}.

\subsubsection{Spectrum ratio}
%'''''''''''''''''''''''''''''
\label{sec:spectrum_ratio}

Consider the stellar flux received by the observer during the
planetary transit to be $F_{\mathrm{in}}$, and the flux
received when the planet is not occulting the star to be
$F_{\mathrm{out}}$. Brown (2001) defined $\Re$ to be the ratio
between those two quantities, and $\Re'$ (the so-called
spectrum ratio) as $\Re'=\Re - 1$. Here, $\Re'$ is the sum of
two distinct types of occultations:
\begin{itemize}
\item The occultation by the `solid' surface of the planet,
optically thick at all wavelengths. Projected along the line of
sight, this is a disk of radius $R_P$ and the occultation depth
is simply $(R_P/R_\star)^2$.
 \item The wavelength-dependent occultation by the thin ring of gaseous components
 that surrounds the planetary disk, which can be expressed as $\Sigma_\lambda /
 (\pi R_\star^2)$. The area, $\Sigma_\lambda$, is the
 atmospheric equivalent surface of absorption and may be calculated as:
\begin{equation}
\Sigma_\lambda = \int_{R_P}^{b_\mathrm{max}} 2 \pi b \mathrm{d}
b \left[1 - \mathrm{e}^{-\tau_\lambda(b)}\right].
\end{equation}
\end{itemize}
The resulting spectrum ratio is:
\begin{equation} \label{eq:spectrum_ratio}
\Re'(\lambda) = - \frac{\Sigma_\lambda + \pi R_P^2}{\pi
R_\star^2}.
\end{equation}
Note that $\Re' < 0$.

\subsection{Description of the atmospheric profiles}
%---------------------------------------------------

Along a single cord, stellar photons are crossing several
levels of the spherically stratified atmosphere. We generate an
atmospheric model using the vertical profiles from Tinetti et
al.\ (2005a, 2005b) and Fishbein et al.\ (2003) for the Earth
and from the Venus International Reference Atmosphere (VIRA,
Kliore et al.~1985) for Venus. These atmospheric data include
the profiles of pressure, $p$, temperature, $T$, and various
mixing ratios, $Y$. The atmospheres are initially sampled in 50
levels, ranging from the ground level to an altitude of about
80~km for the Earth and  about 50~km for Venus. Both profiles
stop below the homopause, so we assume hydrostatic equilibrium
for the vertical pressure gradient.

A useful quantity to describe atmospheres in hydrostatic
equilibrium is the scale height, $H$, i.e. the height above
which the pressure decreases by a factor $e$. The scale height
explicitly depends on the temperature, as $H = k \mathcal{N}_A
T / (\mu g)$, where $k$ and $\mathcal{N}_A$ are the Boltzmann's
and Avogadro's constants while $\mu$ is the mean molar mass of
the atmospheric gas. Since $g$ is the acceleration due to
gravity, $H$ also implicitly depends on the radius and the
density of the planet\footnote{To avoid confusion between the
density of the atmosphere and the mean density of the planet,
the latter is denoted $\rho_P$}. Consequently, less dense
objects are likely to have more extensive atmospheres, hence
they are easier to detect (Brown 2001).

Density and size of planets are therefore key parameters for
the present work. In order to estimate their influence, we test
a set of different planetary types ranging from the Titan-like
giant planet's satellite ($\rho_P \approx 2$~g\,cm$^{-3}$, $R_P
\approx 0.5$~Earth radius -- 0.5~R$_\oplus$) to the
`super-Earth' object ($\rho_P \approx 6$~g\,cm$^{-3}$, $R_P
\approx 2$~R$_\oplus$). For the physical properties of
plausible, theoretically predicted planets such as a
`super-Earth', we use the mass-radius relation model from
Dubois (2004) and from Sotin et al.\ (2005). Our atmospheric
model allows the re-scaling of vertical profiles depending on
the acceleration due to gravity of the planet and the
atmospheric pressure at the reference level.

\subsubsection{Molecular composition of the atmosphere}
%''''''''''''''''''''''''''''''''''''''''''''''''''''''

Our simplified atmospheric profiles contain only the species
that may produce interesting spectral signatures in the chosen
wavelength range (0.2 to 2~$\mu$m), viz., water vapor (H$_2$O),
carbon dioxide (CO$_2$), ozone (O$_3$) and molecular oxygen
(O$_2$). Molecular nitrogen (N$_2$) has also been considered,
though lacking marked electronic transitions from the UV to the
near IR. Nevertheless, it is a major species in Earth's
atmosphere and it has a detectable signature via Rayleigh
scattering at short wavelengths.

We consider three types of atmospheres: (A) N$_2$/O$_2$-rich,
(B) CO$_2$-rich and (C) N$_2$/H$_2$O-rich cases. The first two
types can be associated with existing planetary atmospheres,
respectively Earth and Venus. The last type (C) could
correspond to the atmosphere of an Earth-mass volatile-rich
planet such as an `ocean-planet' described by L\'eger et al.\
(2004). The basis for building a `toy model' of an H$_2$O-rich
atmosphere are found in L\'eger et al.\ (2004) and Ehrenreich
et al.\ (2005b, see Sect.~\ref{sec:H2O-rich_atmo}).

Vertical gradients in the chemical composition and temperature
of each of these atmospheres are plotted in
Fig.~\ref{fig:A_profile} (N$_2$/O$_2$-rich),
Fig.~\ref{fig:B_profile} (CO$_2$-rich) and
Fig.~\ref{fig:C_profile} (N$_2$/H$_2$O-rich).
Table~\ref{tab:composition} summarizes the mean chemical
compositions of these model atmospheres.

\begin{table*}
\centering
\begin{tabular}{*{8}{c}}
\hline \hline
Type              & $\mu$ (g\,mol$^{-1}$) & $Y_{\mathrm{N}_2} (\%)$ & $Y_{\mathrm{H}_2\mathrm{O}}$ (\%) & $Y_{\mathrm{CO}_2}$ (\%) & $Y_{\mathrm{O}_2}$ (\%) & $Y_{\mathrm{O}_3}$ (\%) & Used for models \\
\hline
N$_2$/O$_2$-rich  & 28.8                               & 78                      & 0.3                               & 0.03                     & 21                      & ${<10^{-3}}^*$ & A1, A2, A3 \\
CO$_2$-rich       & 43.3                               & 4                       & $3\cdot10^{-4}$                   & 95                       & 0                       & 0          & B1, B2, B3 \\
N$_2$/H$_2$O-rich & 28.7                               & 80                      & 10                                & 10                       & 0                       & 0          & C1, C2, C3 \\
\hline
\end{tabular}
\caption{Mean volume mixing ratio of atmospheric absorbers for
the different types of model atmospheres considered.
\newline
(*) Ozone is only present in model~A1.}
\label{tab:composition}
\end{table*}

\subsubsection{Temperature profiles}
%'''''''''''''''''''''''''''''''''''
\label{sec:temperature_profile}

As mentioned above, we use Earth and Venus vertical temperature
profiles as prototype for N$_2$/O$_2$-rich and CO$_2$-rich
atmospheres (see Sect.~\ref{sec:choice}). Moreover we assume an
isothermal profile in the thermosphere, instead of the real
one. This is an arbitrary, but conservative choice, since the
temperature should on the contrary rise in the thermosphere
enhancing the atmosphere's detectability (see
Sect.~\ref{sec:temperature_effect}).

\subsubsection{Upper limit of the atmosphere}
%''''''''''''''''''''''''''''''''''''''''''''
\label{sec:b_max}

We set the profiles to extend up to a critical height
$b_\mathrm{max}$ from the centre of the planet, or
$h_\mathrm{max}$ from the surface (\mbox{$b_\mathrm{max} = R_P
+ h_\mathrm{max}$}). This limit corresponds to the altitude
above which the molecular species we considered (H$_2$O, O$_3$,
CO$_2$, O$_2$) are likely to be destroyed or modified either by
photo-dissociating or ionizing radiations, such as Ly$_\alpha$
or extreme-UV (EUV).

Therefore, the critical height corresponds to the mesopause on
Earth (at $\approx 85$~km). The column density of the
terrestrial atmosphere above that altitude, $\mathcal{N}_{\geq
85\mathrm{\,km}}$, is sufficient to absorb all Ly$_\alpha$
flux. In fact, as the number density of the atmospheric gas,
$n(h)$, decreases exponentially with height, we can simply
consider \mbox{$\mathcal{N}_{\geq 85\mathrm{\,km}} \propto
n_{85 \mathrm{\,km}} \cdot H_{85 \mathrm{\,km}}$}, where $n_{85
\mathrm{\,km}}$ and $H_{85\mathrm{~km}}$ are the density and
the scale height of the terrestrial atmosphere at 85~km,
respectively.

Similarly, we set the upper limit of a given atmosphere,
$h_\mathrm{max}$, to the altitude below which the
photo-dissociating photons are absorbed. We assume that
$h_\mathrm{max}$ is the altitude where the column density
equals that of the terrestrial atmosphere at 85~km, that is
\mbox{$n_{h_\mathrm{max}} \cdot H_{h_\mathrm{max}} = (n_{85
\mathrm{\,km}})_\oplus \cdot (H_{85 \mathrm{\,km}})_\oplus$}.
We determine $h_\mathrm{max}$ by scaling this equation.

Values of $h_\mathrm{max}$ for the different models are given
in Table~\ref{tab:models}. Similarly to neutral elements
absorbing light below $h_\mathrm{max}$, it is likely that
ionized elements are absorbing light above this limit, though
we do not include this effect in the model.

\begin{figure}
\resizebox{\hsize}{!}{\includegraphics{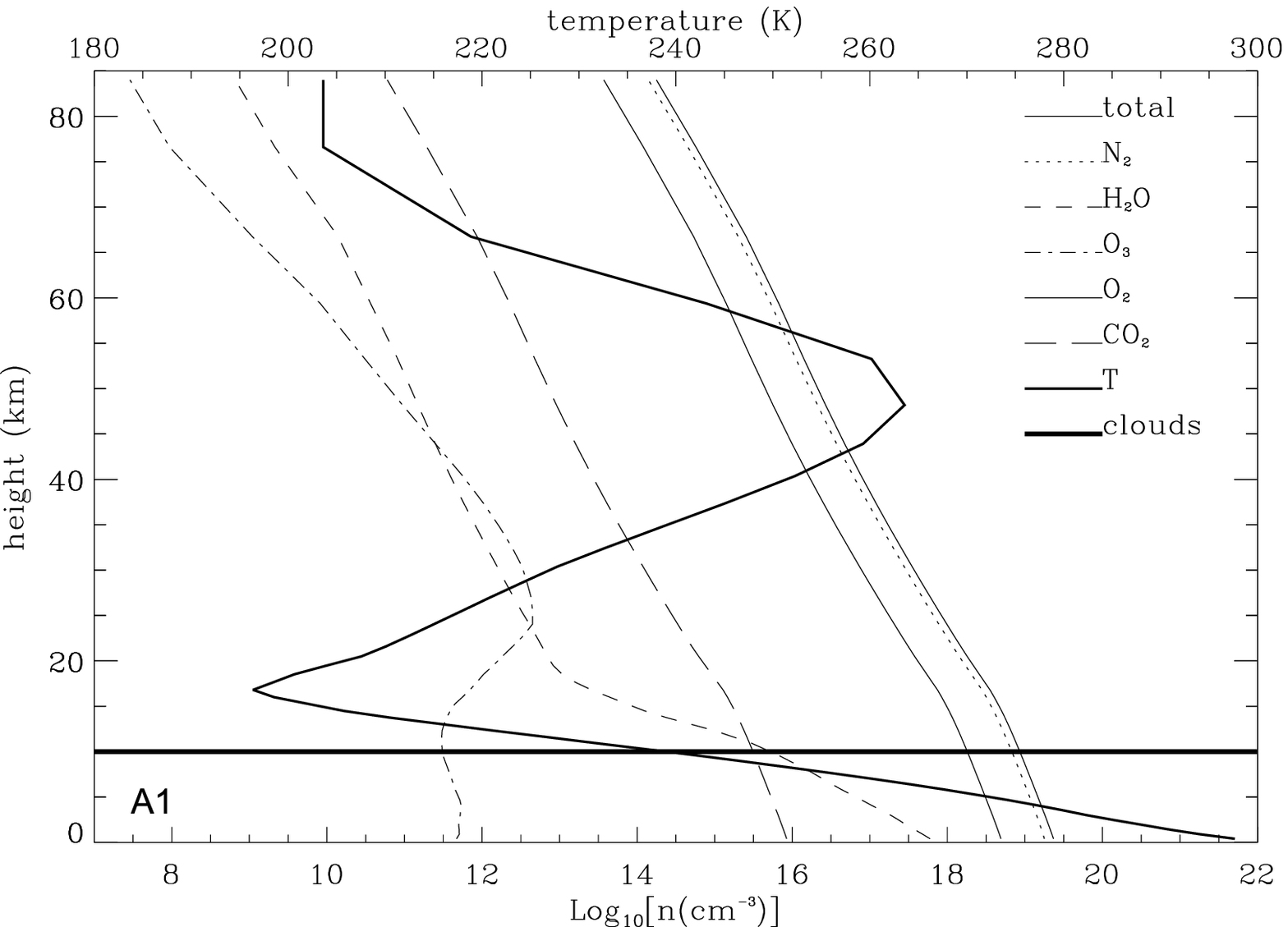}}
\caption{Atmospheric profiles, A1. The plot shows the total
number density profile (thin solid line) of the atmosphere in
cm$^{-3}$, and that of the five species included in our model,
namely, N$_2$ (dotted line), O$_2$(dash-dot-dot-dotted line),
H$_2$O (dashed line), CO$_2$ (long-dashed line) and O$_3$
(dash-dotted line). Temperature (thick line up to 80~km) and
mixing ratios of the different species are those of Earth.
Temperature is assumed to be constant above that height. The
thickest horizontal line shows the position of the cloud
layer.} \label{fig:A_profile}
\end{figure}

\begin{figure}
\resizebox{\hsize}{!}{\includegraphics{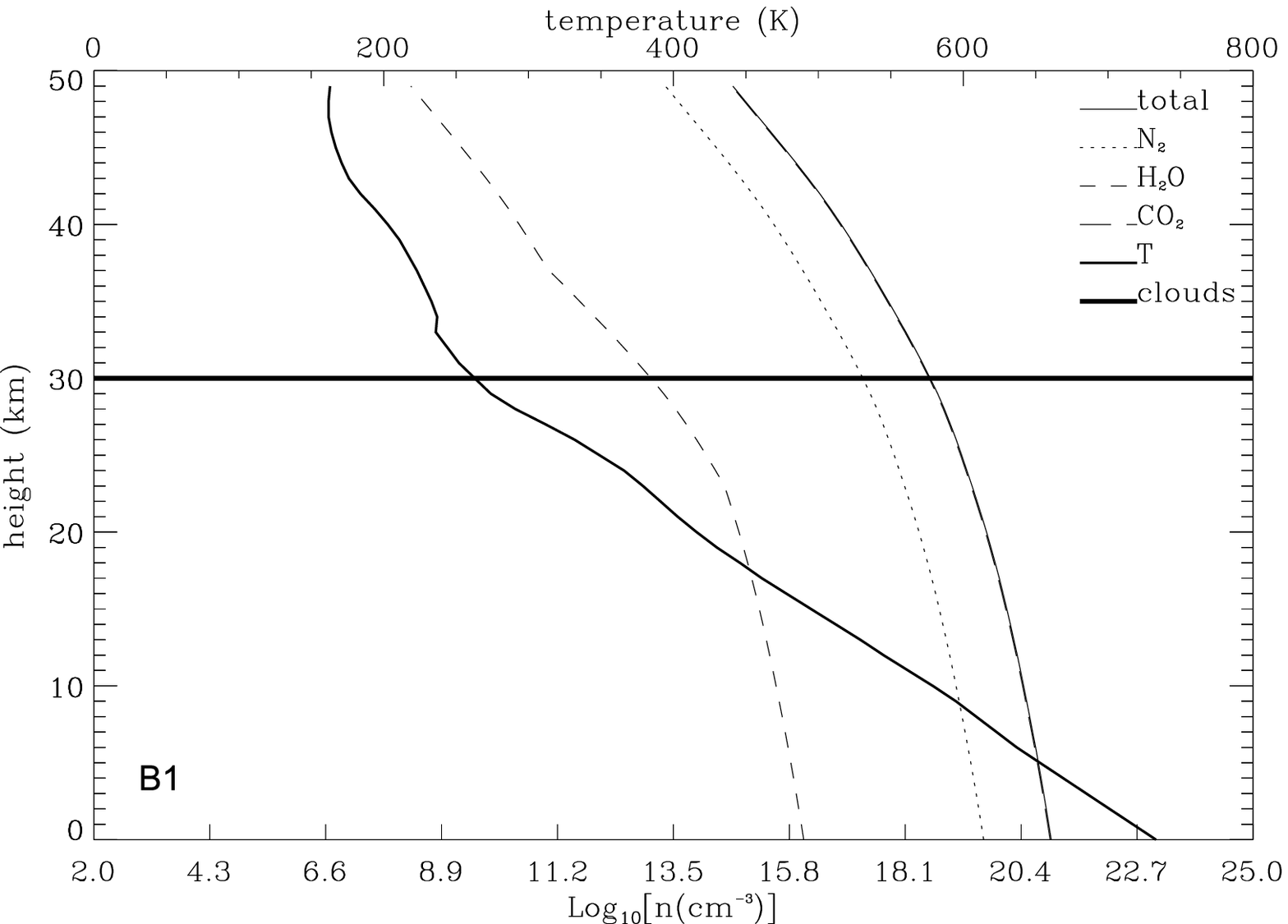}}
\caption{Atmospheric profiles, B1. The legend is identical to
that in Fig.~\ref{fig:A_profile}. The temperature profile and
mixing ratios are that of Venus. The temperature is considered
to be constant above 50~km. Carbon dioxide is barely visible
because it is by far the major constituent so its line is
superimposed with that of the total density.}
\label{fig:B_profile}
\end{figure}

\begin{figure}
\resizebox{\hsize}{!}{\includegraphics{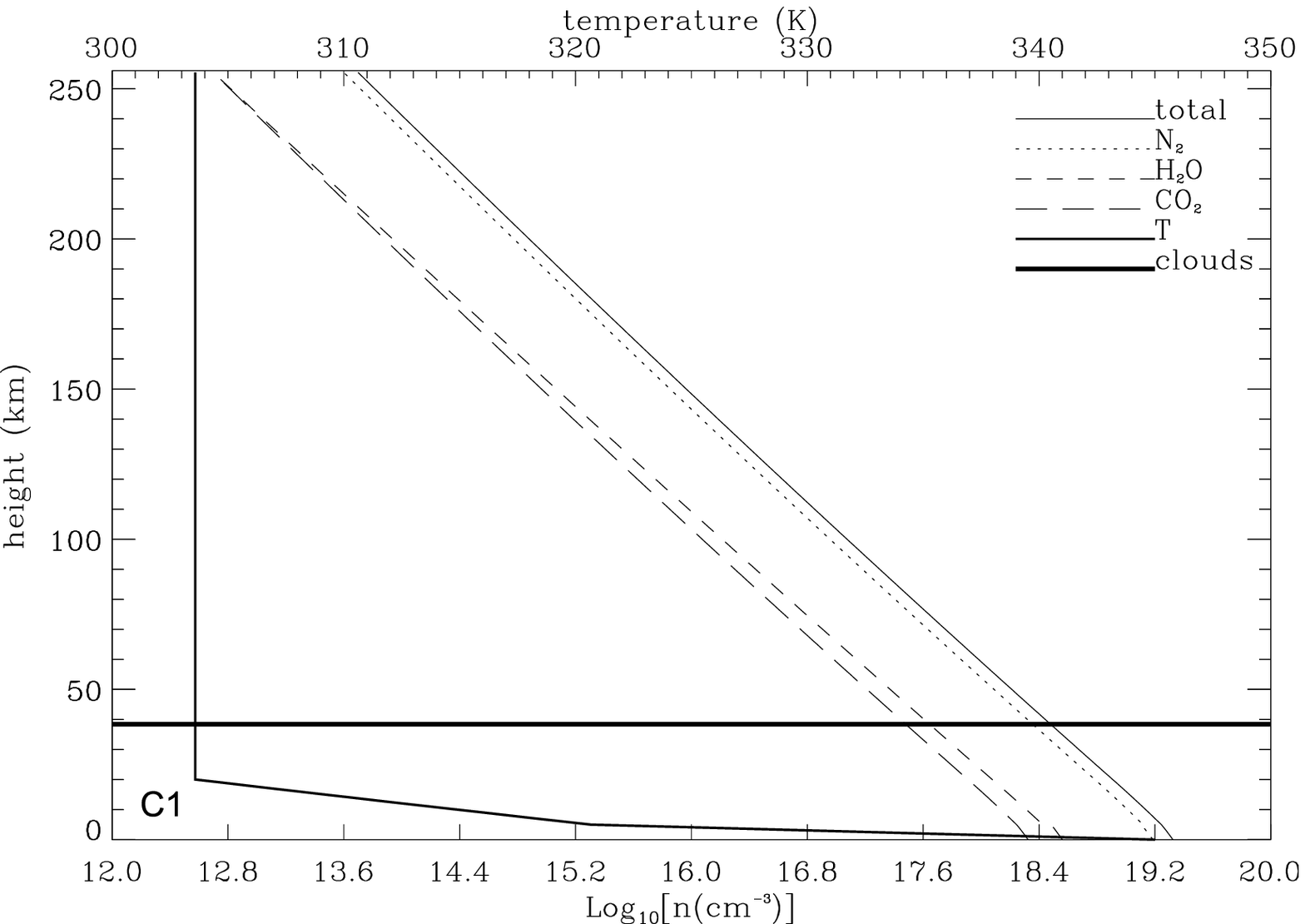}}
\caption{Atmospheric profiles, C1. Same legend as in
Fig.~\ref{fig:A_profile} and Fig.~\ref{fig:B_profile}. The
temperature profile follows a dry adiabat in the first 10~km of
the atmosphere, until the point where $e \geq e_\mathrm{sat}$.
Next, it follows a steeper saturated adiabat up to 20~km high.
The temperature gradient is arbitrarily set to be isothermal
above this point. The cloud top (thickest line) is one scale
height above the higher point where $e \geq e_\mathrm{sat}$.
For reasons detailed in the text (see
Sect.~\ref{sec:H2O-rich_atmo}), this point corresponds to the
level where the temperature gradient becomes isothermal.}
\label{fig:C_profile}
\end{figure}

\subsubsection{Presence of clouds}
%'''''''''''''''''''''''''''''''''
\label{sec:clouds}

In the wavelength range of interest, the surface of Venus is
almost completely hidden by clouds. Therefore, it seems
reasonable to model these types of clouds to a first order
approximation by assuming that they act as an optically thick
layer at a given altitude. As a result, clouds effectively
increase the apparent radius of the planet and the transiting
spectrum gives information only about atmospheric components
existing above the cloud layer. The top of the cloud layer is a
free parameter for N$_2$/O$_2$- and CO$_2$-rich atmospheres
(set to 10 and 30~km, taken from the Earth and Venus,
respectively). We treat the case of the N$_2$/H$_2$O-rich
atmosphere separately because H$_2$O is a highly condensable
species.

\subsubsection{Composition, vertical structure and location of the clouds
in a N$_2$/H$_2$O-rich atmosphere} %'''''''''''''''''''''''''''''''''''''
\label{sec:H2O-rich_atmo}

The temperature gradient of an atmosphere containing
non-negligible amount of condensable species, like H$_2$O,
significantly departs from the case where no condensation
occurs. A correct estimation of the temperature profile is
crucial to determine the scale height, hence the detectability
of that atmosphere. In an H$_2$O-rich atmosphere, the evolution
of the adiabatic temperature gradient is driven by the ratio of
the partial pressure of water vapor, $e$, to the saturating
vapor pressure, $e_\mathrm{sat}$. This ratio should also
determine the levels at which the water vapor is in excess in
the air and condenses (for $e / e_\mathrm{sat}
> 1$), i.e.\ the levels where clouds may form.

Our initial conditions at the $z=0$ level ($z^0$) are the
temperature $T^0$ and pressure $p^0$. With these quantities we
can estimate $e_\mathrm{sat}$, which depends only on the
temperature, using the Clausius-Clapeyron equation:
\begin{equation} \label{eq:Clausius-Clapeyron}
e_\mathrm{sat}(T) = p^*
\exp{\left[\frac{\mu_{\mathrm{H}_2\mathrm{O}}
L_v}{\mathcal{N}_A k} \left( \frac{1}{T^*} - \frac{1}{T}
\right) \right]}
\end{equation}
where $p^*$ and $T^*$ are the reference pressure
($1.013\cdot10^{5}$~Pa) and temperature (373~K),
$\mu_{\mathrm{H}_2\mathrm{O}}$ is the molar mass of water and
$L_v$ is the latent heat of vaporization for water
($2.26\cdot10^{10}$~erg\,g$^{-1}$). Assuming that the planet is
covered with liquid water (e.g., an ocean-planet; see L\'eger
et al.\ 2004) and that $T^0$ is `tropical' (e.g. 340~K), the
humidity at the surface is high so that the value of $e^0$ must
be an important fraction of $e_\mathrm{sat}(T^0)$. We set $e^0$
to half the value of $e_\mathrm{sat}(T^0)$. The volume mixing
ratio of water can be expressed as $Y_{\mathrm{H}_2\mathrm{O}}
= e / p$, and we can calculate it at the surface of the planet.
The atmosphere of an ocean-planet may also contain a
significant quantity of CO$_2$. We arbitrarily set this
quantity constant to $Y_{\mathrm{CO}_2} = 0.1$ (L\'eger et al.\
2004; Ehrenreich et al.\ 2005b). Molecular nitrogen is the
major constituent of the atmosphere of the Earth and the second
more abundant species in the atmosphere of Venus, and therefore
we chose to include it to complete the chemical composition of
this atmosphere. The mixing ratio of N$_2$ was set to be
$Y_{\mathrm{N}_2} = 1 - Y_{\mathrm{CO}_2} -
Y_{\mathrm{H}_2\mathrm{O}}$ at any level. Assuming the
atmosphere contains only N$_2$, H$_2$O and CO$_2$, we can
obtain the mean molar mass of the atmospheric gas ($\mu^0 =
\sum_i Y_i^0 \mu_i$) and that of the dry atmospheric gas
($\mu_\mathrm{d}^0 = \mu^0 - Y_{\mathrm{H}_2\mathrm{O}}^0
\mu_{\mathrm{H}_2\mathrm{O}}$), the mean specific heat of dry
air ($C_p^0 = \sum {C_p}_i Y_i^0 \mu_i / \mu_\mathrm{d}^0$) and
the scale height $H_0$ (all at the level $z^0$).

For the $z^{j+1}$ level, we need to evaluate the temperature
gradient between $z^j$ and $z^{j+1}$. There are two cases
(Triplet \& Roche 1986):
\begin{itemize}
\item $e^j < e_\mathrm{sat}^j$; in this case the temperature
follows a dry adiabatic gradient,
\begin{equation} \label{eq:dry_gradient}
{\Delta T}_\mathrm{dry} = \frac{-g}{C_p^j}.
\end{equation}
\item  $e^j = e_\mathrm{sat}^j$; in this case the gradient is
saturated,
\begin{equation}
\label{eq:sat_gradient} {\Delta T}_\mathrm{sat} = {\Delta
T}_\mathrm{dry} \frac{\left( 1 + r_\mathrm{sat}^j \right)
\left[ 1 + L_v r_\mathrm{sat}^j / (R_\mathrm{dry}^j T^j)
\right]}{1 + \frac{r_\mathrm{sat}^j}{C_p^j}
\left[{C_p}_{\mathrm{H}_2\mathrm{O}} + L_v^2 \frac{1 +
r_\mathrm{sat}^j
R_{\mathrm{H}_2\mathrm{O}}R_\mathrm{dry}^j}{R_{\mathrm{H}_2\mathrm{O}}
(T^j)^2} \right]}
\end{equation}
where $r_\mathrm{sat}^j = (\mu_{\mathrm{H}_2\mathrm{O}}
e_\mathrm{sat}^j) / [\mu_\mathrm{d}^j (p^j - e_\mathrm{sat}^j)
]$ is the mixing ratio of saturated air, $R_\mathrm{dry}^j =
\mathcal{N}_A k / \mu_\mathrm{dry}^j$ and
$R_{\mathrm{H}_2\mathrm{O}} = \mathcal{N}_A k /
\mu_{\mathrm{H}_2\mathrm{O}}$ are the specific constant of dry
air at the level $z^j$ and water (respectively).
\end{itemize}
If $z^{j+1} < 20$~km, we select the appropriate gradient
accordingly to the value of $e / e_\mathrm{sat}$, and get the
value of the temperature $T^{j+1}$. Above 20~km, we assume the
temperature profile becomes isothermal ($T^{j+1} = T^j$).

The assumption of an isothermal atmosphere, already discussed
in Sect.~\ref{sec:temperature_profile}, is somewhat arbitrary
but is motivated by an analogy with the atmosphere of the
Earth, where the temperature gradient becomes positive from
about 20 to 50~km. Taking an isothermal temperature gradient
will conservatively mimic the presence of a stratosphere.
However, it has important consequences since it allows H$_2$O
to be significantly present above the cloud top. In fact, above
20~km, the temperature stops decreasing, preventing
condensation from occurring (the saturation vapor pressure
depends only on temperature). Our assumption consequently fixes
the height of the cloud deck to the point where the temperature
profile is isothermal (actually, one scale height above that
point). If we set this point higher, we would increase the
amount of clouds hence reducing the detectable portion of
atmosphere. In addition, the cloud formation would certainly
take the corresponding latent heat of condensation out of the
atmospheric gas, contributing, as a consequence, to cool the
atmosphere at the level of the cloud layer.

We calculate $H^{j+1}$, $p^{j+1} = p^j \cdot
\exp{\left(-z^{j+1}/H^{j+1}\right)}$, $e_\mathrm{sat}^j$ (from
Eq.~\ref{eq:Clausius-Clapeyron}) and either $e^{j+1} = e^j
\cdot \exp{\left[(z^j - z^{j+1}) / H^{j+1}) \right]}$, if the
atmosphere is not saturated or $e^{j+1} =
e_\mathrm{sat}^{j+1}$, if the atmosphere is saturated. We
finally find all $Y_i^{j+1}$, $\mu_\mathrm{dry}^{j+1}$ and
${C_p}_\mathrm{dry}^{j+1}$ and then iterate the process for all
atmospheric levels.

The higher and the lower pressure levels where $e =
e_\mathrm{sat}$ indicate respectively the bottom and the top of
the region where clouds are forming. We assume the cloud layer
does not extend over one scale height above the top of the
cloud forming region. However, we can still have $e \leq
e_\mathrm{sat}$ higher in the atmosphere, and thus H$_2$O can
be present above the clouds.

\subsection{Description of atmospheric absorptions}
%--------------------------------------------------

\subsubsection{Chemical species}
%'''''''''''''''''''''''''''''''
We used the program \texttt{LBLABC} (Meadows \& Crisp 1996), a
line-by-line model that generates monochromatic gas absorption
coefficients from molecular line lists, for each of the gases,
except ozone, present in the atmosphere. The line lists are
extracted from the HITRAN~2000 databank (Rothman et al.\ 2003).
We calculated the absorption coefficients for O$_2$, H$_2$O and
CO$_2$ in our wavelength range we (i.e., from 200 to
2\,000~nm).

\begin{figure}
\resizebox{\hsize}{!}{\includegraphics{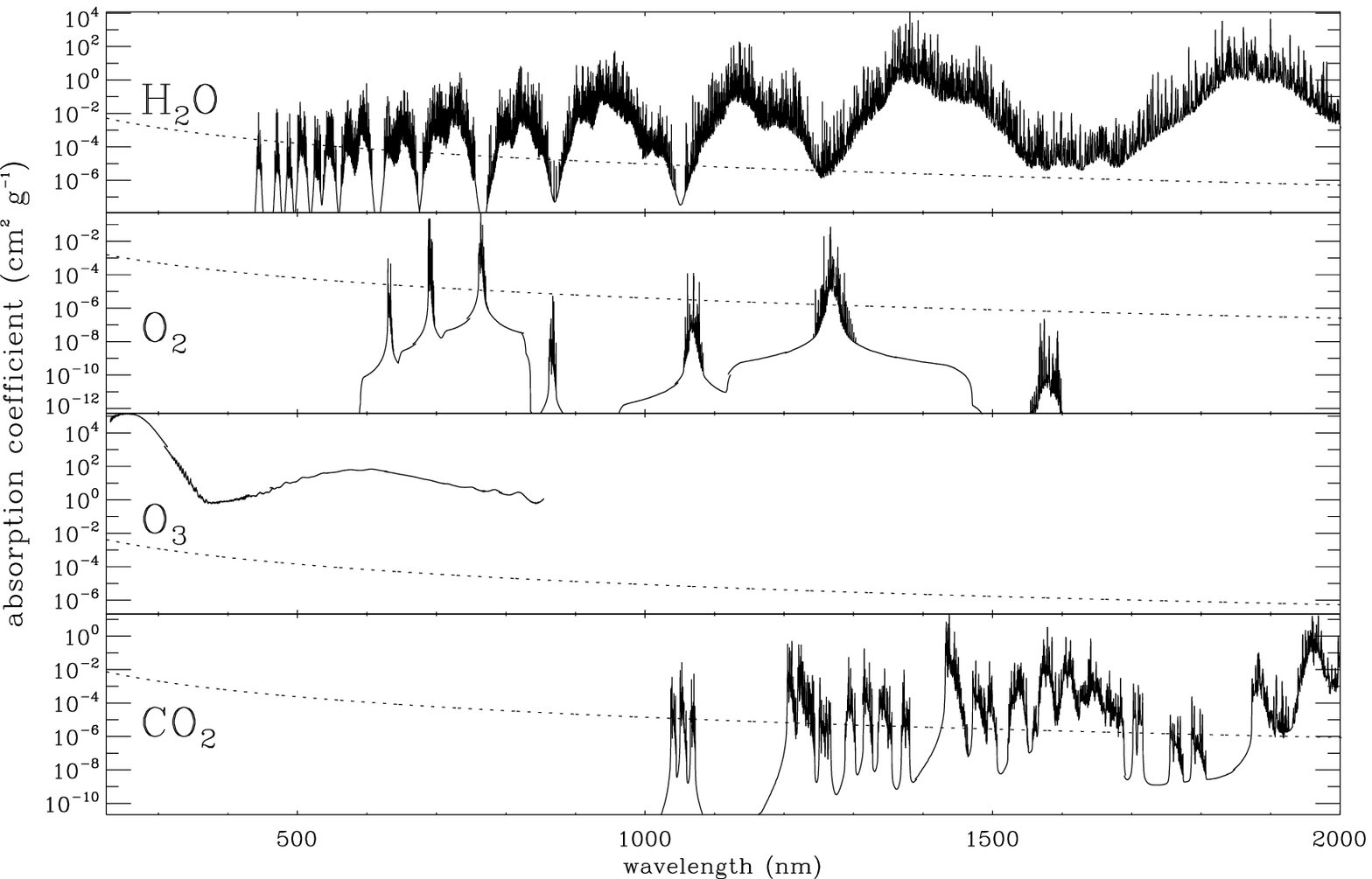}}
\caption{Absorption coefficients of atmospheric absorbers (in
cm$^2$\,g$^{-1}$), as a function of the wavelength. The
photo-absorption coefficients corresponding to H$_2$O, O$_2$,
O$_3$ and CO$_2$ (solid lines) are plotted against their
respective Rayleigh scattering coefficient (dotted line),
except O$_3$, plotted against the Rayleigh scattering
coefficient of N$_2$.} \label{fig:abc}
\end{figure}

The absorption coefficients relative to these species depend on
pressure and temperature. We verified that those variations do
not impact significantly on the results obtained (see
Sect.~\ref{sec:results}) and we decided to use the absorption
coefficients calculated at the pressure and temperature of the
cloud layer, i.e., 10~km in models~A1, A2 \&~A3, 30~km in
models~B1, B2 \&~B3 and from 25 to 70~km in models~C1 to~C3. We
then assumed these absorption coefficients to be constant along
the $z$-axis. This is a fairly good approximation since
molecules at that atmospheric level contribute more
substantially to the transmitted spectrum than molecules at the
bottom of the atmosphere. Absorption coefficients for H$_2$O,
CO$_2$, O$_3$ and O$_2$ are compared in Fig.~\ref{fig:abc}.

The spectrum of O$_3$ is unavailable in HITRAN at wavelengths
lower than 2.4~$\mu$m. However it has strong absorption in the
Hartley (200--350~nm) and Chappuis (400--750~nm) bands. Thus we
took the photo-absorption cross-sections, $\sigma$ (in
cm$^{2}$), from the GEISA/cross-sectional databank
(Jacquinet-Husson et al.\ 1999) and converted them into
absorption coefficients, $A$ (in cm$^{2}$\,g$^{-1}$), such as
$A = \sigma \mathcal{N}_A / \mu$, where $\mu$ is the molar mass
of the component.

\begin{figure}
\resizebox{\hsize}{!}{\includegraphics{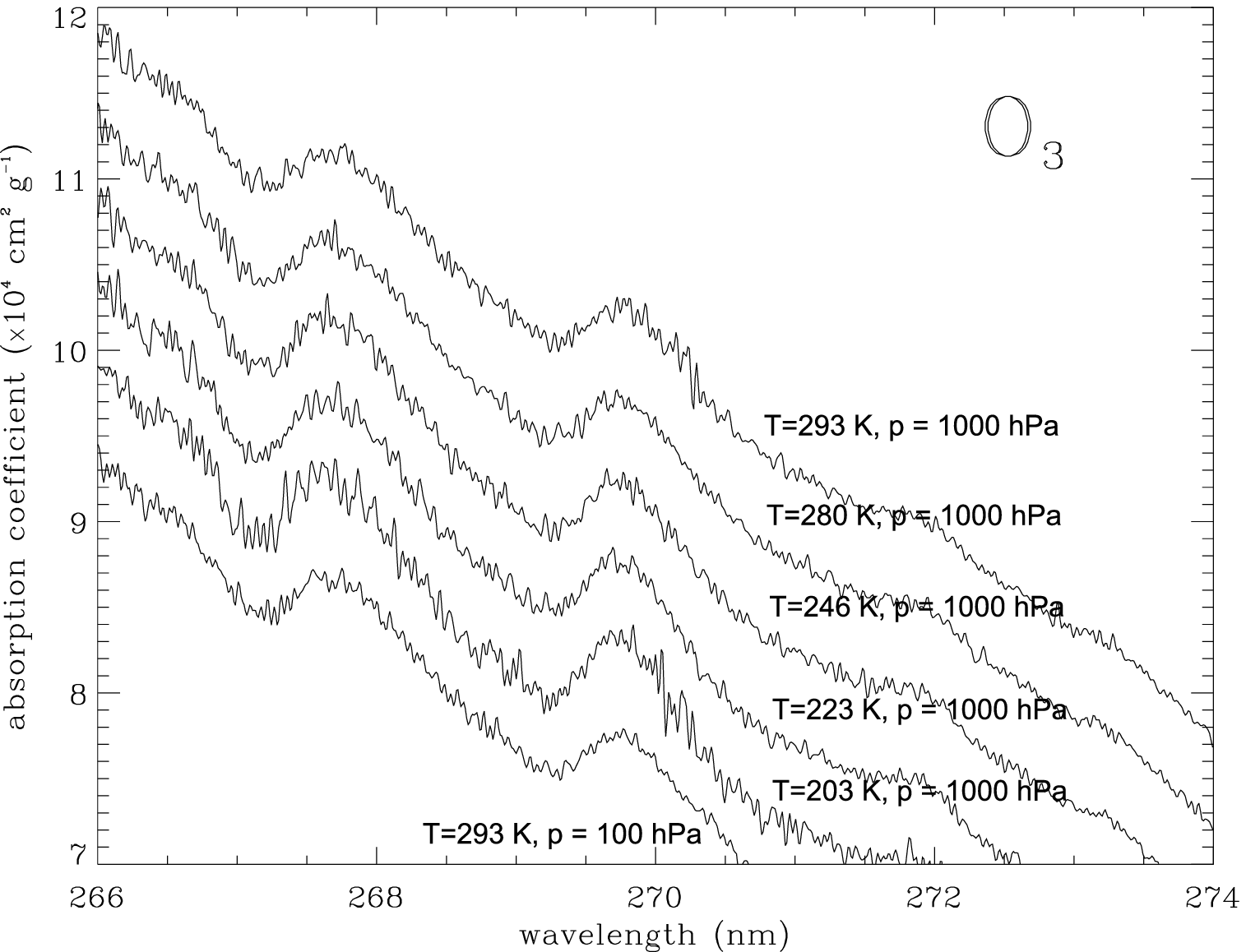}}
\caption{Dependence of the absorption coefficient of O$_3$ on
pressure and temperature. For clarity, each line has been
shifted down by $5 \cdot 10^4$~cm$^2$\,g$^{-1}$ with respect to
the previous one.} \label{fig:abc_variation}
\end{figure}

As shown in Fig.~\ref{fig:abc_variation}, the pressure and the
temperature variations do not have a significant influence over
the cross sections/absorption coefficients of O$_3$. We
therefore used the values given for $p = 1$~atm\footnote{1~atm
= 1\,013~hPa.} and $T = 300$~K, and set them constant along the
$z$-axis.

\subsubsection{Rayleigh diffusion}
%'''''''''''''''''''''''''''''''''
Light is scattered toward short wavelengths by atmospheric
molecules whose dimensions are comparable to $\lambda$.
Rayleigh diffusion could be an important indicator of the most
abundant atmospheric species. Molecular nitrogen, for instance,
does not present any noticeable spectroscopic lines between 0.2
and 2~$\mu$m. With a transit observation, the presence of a gas
without spectroscopic lines like nitrogen in the Earth
atmosphere can be indirectly inferred from the
wavelength-dependance of the spectrum ratio continuum. Since
Rayleigh scattering cross section of CO$_2$ is high, Venus-like
atmospheric signatures should also present an important
Rayleigh scattering contribution.

We have therefore estimated these different contributions. The
Rayleigh scattering cross section, $\sigma_R$, can be expressed
in cgs units as: (Bates 1984; Naus \& Ubachs 1999; Sneep \&
Ubachs 2004)
\begin{equation} \label{eq:rayleigh_xsc}
\sigma_R(\bar{\nu}) = \frac{24 \pi^3 \bar{\nu}^4}{n^2} \left(
\frac{r(\bar{\nu})^2 - 1}{r(\bar{\nu})^2 + 2} \right)
\end{equation}
where $\bar{\nu} = 1 / \lambda$, $n$ is the number density
(cm$^{-3}$) and $r$ is the refractive index of the gas. The
total Rayleigh scattering includes weighted contributions from
N$_2$, O$_2$, CO$_2$ and H$_2$O (i.e., $\sigma_R = \sum_{i} Y_i
{\sigma_R}_i$), and so we need all the corresponding refractive
indexes. These are found in Bates (1984) and Sneep \& Ubachs
(2004) for N$_2$, O$_2$ and CO$_2$.\footnote{We noted a
typographical error in the CO$_2$ refractive index formula
(Eq.~13) in Sneep \& Ubachs (2004): in order to yield the
correct values, results from this expression should be divided
by $10^3$ (M.~Sneep, personal communication).} The refractive
index for H$_2$O comes from Schiebener et al.\ (1990). Tests
have proved the different refractive indexes do not
significantly change with temperature and pressure. We have
therefore calculated the indexes for standard conditions
($15\degr$C and 1013~hPa).

\subsubsection{Refraction}
%'''''''''''''''''''''''''
Depending on the wavelength, the refraction may bring into the
line of sight rays coming from different parts of the star. To
quantify the importance of that effect, we calculate the
maximum deviation, $\Delta\theta$, due to the wavelength
dependence of the refraction index, using the formula given by
Seager \& Sasselov (2000) and the refractive index at the
surface ($h = 0$) between 0.2 and 2~$\mu$m. We obtain
\mbox{$\Delta\theta \approx 0.3\arcmin$}. This represents about
1.5\%, 1\% and 0.5\% of the angular diameter of the star (F-,
G- and K-type star, respectively) as seen from the planet. We
can therefore consider this effect negligible as long as there
are no important variations of the stellar flux on scales lower
than the surface corresponding to these numbers.

\subsection{Choice of test models}
%---------------------------------
\label{sec:choice}

We chose 9 cases, divided into 3 categories:
1~R$_\oplus$-planets (models~A1, B1 and~C1),
0.5~R$_\oplus$-planets (A2, B2 and~C2) and 2~R$_\oplus$-planets
(A3, B3 and~C3). The parameters for each model are summarized
in Table~\ref{tab:models}. For theses ranges of planetary radii,
the depth of the occultation by the tested planets will differ
by a factor of $\sim$16 at most during their transit. Notice
that a better detection of the transit itself does not always
imply a better detection for the atmosphere of the transiting
planet. On the contrary, in some cases, the fainter the transit
is, the more detectable the atmosphere will be!
In any case, we naturally need to secure the detection of the planet
itself before looking for an atmosphere.

The choice of studying planets with a variety of sizes gives us
the possibility to explore a large range of planet
characteristics, in mass, radius and density. The Earth density
is 5.5~g\,cm$^{-3}$. A planet having the internal composition
of the Earth and twice its radius would weigh $\sim$10 times
more, while a planet half large would weigh $\sim$10 times less
(Sotin et al.~2005). That gives densities of 6.1 and
4~g\,cm$^{-3}$, respectively. We thus have 3 cases, each of
which can be coupled with a plausible atmosphere. We chose a
N$_2$/O$_2$-rich atmosphere (similar to that of the Earth) for
models~A1, A2 and~A3, and a Cytherean (i.e.,
Venus-like\footnote{Cythera
(\emph{K}$\acute{\upsilon}\theta\eta\rho\alpha$) is an Ionian
island where, according to the Greek mythology, the goddess
Aphrodite/Venus first set foot. See
\texttt{http://en.wikipedia.org/wiki/Cytherean}.}) CO$_2$-rich
atmosphere for models~B1, B2 and~B3.

Note that the atmospheric pressure profiles are scaled from the
1~R$_\oplus$ cases (A1 and B1) to the 0.5 and 2~R$_\oplus$
models. In doing so, we did not include any species that showed
a peak of concentration in altitude, such as the O$_3$ layer in
model~A1. In fact, the O$_3$ peak does not depend only on the
hydrostatic equilibrium, but also on the photochemical
equilibrium at the tropopause of the Earth. For that reason
O$_3$ is absent in models~A2 and~A3.

L\'eger et al.\ (2004) suggested the existence of
`ocean-planets', whose internal content in volatiles (H$_2$O)
might be as high as 50\% in mass. Such planets would be much
less dense than telluric ones. We are particularly interested
in those ocean-planets since the lower the density of the
planet is, the higher the atmosphere extends above the surface.
These objects could have densities of 1.8, 2.8 and
4.1~g\,cm$^{-3}$ for radii of 0.5, 1 and 2~R$_\oplus$ (Sotin et
al.~2005), which are relatively small, but reasonable if
compared with Titan's density (1.88~g\,cm$^{-3}$). The huge
quantity of water on the surface of an ocean-planet could
produce a substantial amount of water vapor in their
atmosphere, if the temperature is high enough. A non-negligible
concentration of CO$_2$ might be present as well in those
atmospheres (Ehrenreich et al.~2005b). Using this information
on ocean-planets, we can simulate three extra cases, namely~C1,
C2 and~C3 (Table~\ref{tab:models}).

\subsection{Choice of different stellar types}
%---------------------------------------------
\label{sec:distance} In this work, we consider planets orbiting
in the habitable zone (HZ) of their parent star. Our
atmospheric models are not in fact a good description for
planets orbiting too close to their parent star. For instance,
the heating of the atmosphere by an extremely close star could
trigger effects like evaporation, invalidating the hydrostatic
equilibrium we assumed (see, for instance, Lecavelier des
Etangs et al.\ 2004; Tian et al.\ 2005). The reduced semi-major
axis $a_r$ of the orbit of all planets we have considered is
defined as:
\begin{equation} \label{eq:a_r}
a_r = a \cdot (L_\star / L_\odot)^{-0.5}.
\end{equation}
We set $a_r = 1$~astronomical unit (AU), so that the planet is
in the HZ of its star.

Here we focus on Earth-size planets orbiting around different
main sequence stars, such as K-, G- and F-type stars, since the
repartition of stellar photons in the spectrum is different
from one spectral type to another. Planets in the HZ of K, G
and F stars, with $a_r = 1$~AU, should have a real semi-major
axis of 0.5, 1 and 2~AU, respectively.

\section{Signal-to-noise ratio for ideal observations}
%=====================================================
\label{sec:S/N} Prior to the atmospheres, we need to detect the
planets themselves with a dedicated survey, as the one proposed
by Catala et al.\ (2005). The transmission spectroscopy we
theoretically study here require the use of a large space
telescope. Hence, we need to quantify the S/N of such
observations to determine the detectability of the atmospheric
signatures for a transiting Earth-size exoplanet. The S/N will
depend on both instrumental and astrophysical parameters.

\subsection{Instrumental requirements}
%-------------------------------------
\label{sec:S/N_instru} The first relevant parameter relative to
the instrumentation is the effective area of the telescope
collecting mirror, $S$, which can be expressed as $S=(\epsilon
D)^2 \pi / 4$. The coefficient $\epsilon^2$ accounts for the
instrumental efficiency and $\epsilon D$ is thus the `effective
diameter' of the mirror. Up to present, all exoplanetary
atmospheric signatures have been detected by the Space
Telescope Imaging Spectrograph (STIS) on board the \emph{Hubble
Space Telescope (HST)}. This instrument, now no longer
operative, was very versatile\footnote{STIS was used for
imagery, spectro-imagery, coronography and low and high
resolution spectroscopy.} and consequently not planned to have
high efficiency. It had a throughput $\epsilon^2 \approx 2\%$
from 200 to 300~nm, and $\epsilon^2 \approx 10\%$ from 350 to
1\,000~nm. As the majority of photons we are interested in is
available in the range from 350 to 1\,000~nm, we reasonably
assume that a modern spectrograph has a mean $\epsilon^2$
significantly greater than 10\% from 200 to 2\,000~nm. Present
day most efficient spectrographs have $\epsilon^2 \approx 25\%$
in the visible, so it seems reasonable to imagine that next
generation spectrographs, specifically designed to achieve high
sensitivity observations, could have throughput of $\epsilon^2
\approx 25\%$, or $\epsilon = 50\%$.

Another parameter linked to the instrument is the spectral
resolution, $\mathcal{R}$. In the following, $\mathcal{R}$ will
be assumed to be about 200, i.e. 10~nm-wide spectral feature
can be resolved.

Finally, it is legitimate to question the ability of the
instrument detectors to discriminate the tenuous ($\sim$
10$^{-6}$) absorption features in the transmitted spectra of
Earth-size planets. In a recent past, sodium was detected at a
precision of 50~parts-per-million (ppm) on a line as thin as
about 1~nm by Charbonneau et al.\ (2002) using STIS. According
to our results (see Sect.~\ref{sec:results}), some absorption
features from Earth-size planet atmospheres show a $\sim$1~ppm
dimming over $\sim$100~nm: the technological improvement
required to fill the gap should not be unachievable. Besides,
since we deal with \emph{relative} measurements -- the
in-transit signal being compared to the out-of-transit one --
there is no need to have detectors with a perfect absolute
calibration. Only a highly stable response over periods of
several hours is required. Nevertheless, instrumental precision
remains a challenging issue whose proper assessment will
require further, detailed studies.

\subsection{Physical constraints on the observation}
\begin{figure}
\resizebox{\hsize}{!}{\includegraphics{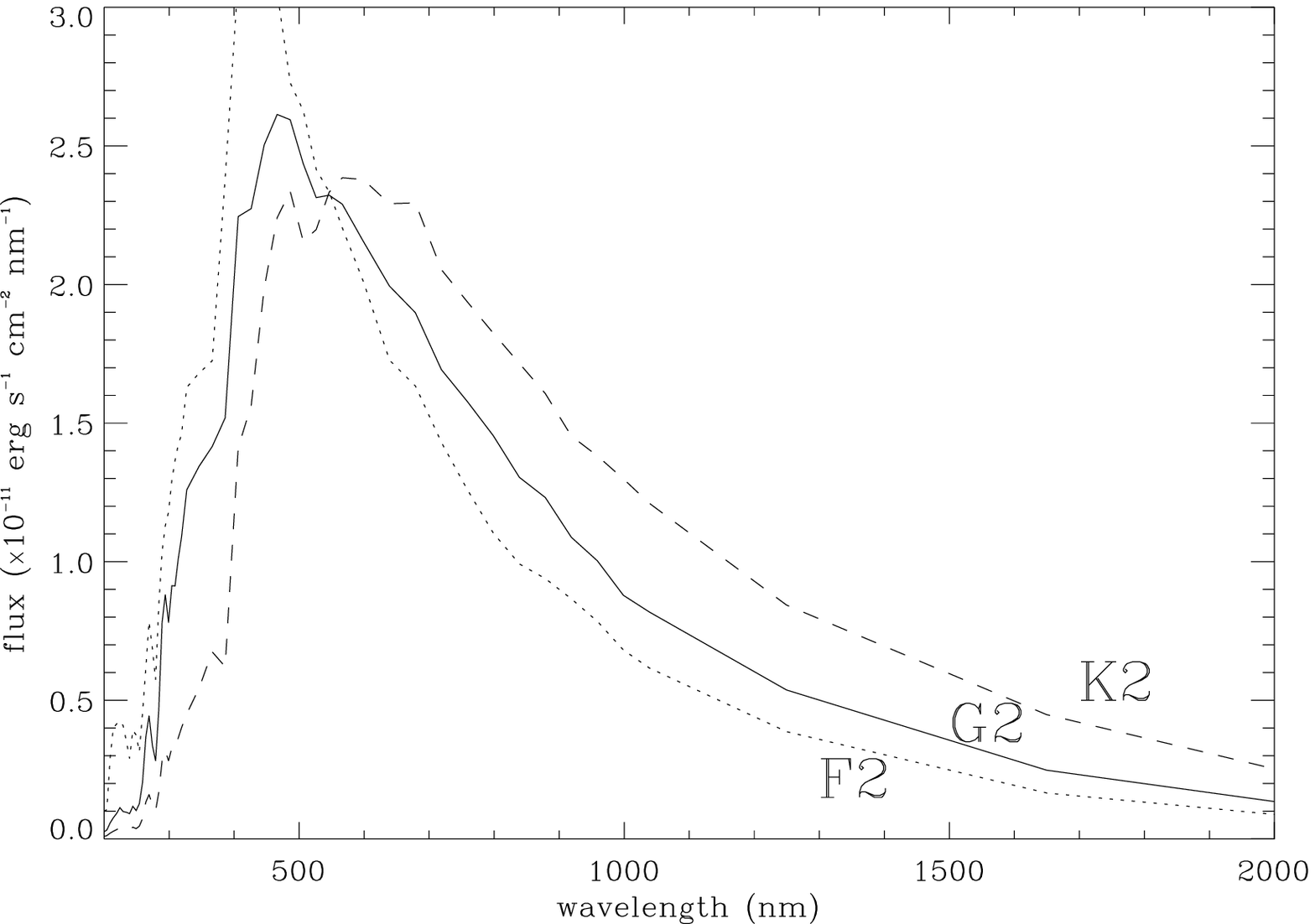}}
\caption{Spectrum of a K2 (dashed line), G2 (solid line), and
F2-type stars (dotted line) between 0.2 and 2~$\mu$m. The
fluxes are scaled to an apparent magnitude $V=8$.}
\label{fig:stars}
\end{figure}

%---------------------------------------------------
\label{sec:S/N_physics} The number of photons detected as a
function of wavelength depends on the spectral type of the
star, while the total number of photons received in an exposure
of duration $t$ depends on the apparent magnitude of the star,
$V$. The stellar spectra $F_{\star}^{V=0}(\lambda)$ are from
\object{$\rho$~Capricorni} (F2\,{\sc iv}), \object{HD~154\,760}
(G2\,{\sc v}) and \object{HD~199\,580} (K2\,{\sc iv}) and are
taken from the Bruzual-Persson-Gunn-Stryker (BPGS)
spectrophotometry atlas\footnote{Available on
\texttt{ftp.stsci.edu/cdbs/cdbs2/grid/bpgs/}.}. The fluxes
(erg\,cm$^{-2}$\,s$^{-1}$\,\AA$^{-1}$) are given at a null
apparent magnitude, so we re-scaled them for any apparent
magnitude $V$, \mbox{$F_\star = F_{\star}^{V=0} \cdot 10^{-0.4
V}$}. The three corresponding spectra are plotted for a default
magnitude $V=8$ in Fig.~\ref{fig:stars}.

The stellar type determines the radius and the mass of the
star, so the transit duration (and thus the maximum time of
exposure during the transit) is different depending on the star
we consider. The transit duration is also a function of the
semi-major axis of the planet orbit. Since we chose a constant
reduced distance ($a_r = 1$~AU) for all planetary models (see
Sect.~\ref{sec:distance}), the duration of transit depends on
the stellar luminosity as well. From Zombeck (1990), we obtain
the radii of F and K stars relatively to that of the Sun,
respectively $R_\mathrm{F}/R_\odot \approx 1.25$ and
$R_\mathrm{K}/R_\odot \approx 0.75$, the mass ratios,
$M_\mathrm{F}/M_\odot \approx 1.75$ and $M_\mathrm{K}/M_\odot
\approx 0.5$, and the luminosity ratios, respectively
$L_\mathrm{F}/L_\odot \approx 4$ and $L_\mathrm{K}/L_\odot
\approx 0.25$. Using Eq.~\ref{eq:a_r}, the duration of the
transit is:
\begin{equation}
\label{eq:time_transit} \tau \approx \frac{13 \pi}{4}~\rm{h}
\cdot \frac{R_\star}{R_\odot}
\left(\frac{M_\star}{M_\odot}\right)^{-0.5}
\left(\frac{L_\star}{L_\odot} \right)^{0.25},
\end{equation}
where $13\pi/4$~h is the mean transit duration of a planet at
1~AU across a G star averaged over all possible impact
parameter of the transit. From Eq.~\ref{eq:time_transit} we
obtain mean transit durations of 7.6, 10.2 and 13.6~h for K-,
G- and F-type star, respectively. In the following, we set $t =
\tau$.

Ideally, our observations are limited only by the stellar
photon noise -- the detection of sodium at a precision of
$\sim$50~ppm in the atmosphere of \object{HD~209\,458b} by
Charbonneau et al. (2002) was in fact limited by the stellar
photon noise. However, at the low signal levels we are
searching for, the intrinsic stellar noise might need to be
considered as well. Stellar activity, as well as convective
motions will cause variations in both intensity and color in
the target stars, on a large variety of timescales. The impact
of stellar micro-variability on the detectability of
photometric transits has been addressed by a number of studies
(see, e.g., Moutou et al., 2005; Aigrain et al., 2004 --
especially their Fig.~8; Lanza et al., 2004), all pointing
towards photometric variability levels in the range of
$\sim$100--1\,000~ppm for durations of a few days. This is to
be compared to the strength and duration of the atmospheric
signatures we want to look at: they are $\sim$1~ppm variations
lasting a few hours. While indeed the different time frequency
and spectral content of these signatures versus the stellar
noise will hopefully allow to discriminate the two, the impact
of stellar micro-variability on such faint signals is likely to
be significant, and may limit the ability to detect an
atmosphere in a transiting planet. For instance, Aigrain et al.
(2004) suggested K stars are more adapted than G or F stars
regarding to the detection of terrestrial planets versus
stellar micro-variability. However, note that the observation
of several transits for each planet considered will confirm the
signal detected in the first transit. For instance, at
$a_r=1$~AU around a K star, a planet has a period of $\approx
0.3$~yr, allowing to schedule several transit observations
within a short period of time. Finally, the usual technique to
detect a spectral signature from a transit is to compare
in-transit and out-of-transit observations (Vidal-Madjar et
al.\ 2003, 2004). For all these reasons, we will assume in the
following to be able to discriminate a transit signal from the
stellar activity and consequently the photon-noise to be the
limiting factor. Nevertheless, further and detailed analysis is
certainly needed to quantify the effect of stellar
micro-variability, as a function of the stellar type, but this
is outside the scope of this paper.

\subsection{Calculation of the signal-to-noise ratio}
%----------------------------------------------------
Now let $\varphi_\star$ be the maximum number of photons per
element of resolution that can be received during $\tau$:
\mbox{$\varphi_\star = F_\star(\lambda) \cdot \lambda / (h_P c)
\cdot \mathcal{R} \cdot S \cdot \tau$}, where $h_P$ is Planck's
constant and $c$ the speed of light. Some photons are blocked
or absorbed by the planet, therefore the actual number of
photons received during the transit is \mbox{$\varphi =
\varphi_\star (1 + \Re')$} per element of resolution.

From the observations, it is possible to obtain $\tilde{R}_P$,
an estimate of the radius of the transiting planet $R_P$ (e.g.,
by using the integrated light curve or a fit to the observed
spectrum ratio). This value corresponds to the flat spectrum
ratio (i.e., a planet without atmosphere) that best fits the
data. The corresponding number of photons received during an
observation per element of resolution is therefore expressed
as: \mbox{$\tilde{\varphi} = \varphi_\star \left[1 -
(\tilde{R}_P / R_\star)^2\right]$}.

The weighted difference between $\varphi$ and $\tilde{\varphi}$
can reveal the presence or the absence of a planetary
atmosphere. We express the $\chi^2$ of this difference over all
the elements of resolution $k$ as \mbox{$\sum_{k} \left[
\left(\varphi_k - \tilde{\varphi}_k \right) /
\sigma_{\varphi_k} \right]^2$}. Here, the uncertainty of the
number of photons received is considered to be dominated by the
stellar photon noise (see Sect.~\ref{sec:S/N_physics}), that is
$\sigma_\varphi = \sqrt{\varphi}$. We thus have:
\begin{equation} \label{eq:chi2}
\chi^2 = \sum_{k} \left( \frac{{\varphi_\star}_k}{1 + \Re'_k }
\left[ \Re'_k + \left(\tilde{R}_P / R_\star \right)^2 \right]^2
\right).
\end{equation}
Given the $\chi^2$, the S/N can be directly calculated taking
its square root. The best estimation can be obtained by
minimizing the $\chi^2$ with respect to the radius
$\tilde{R}_P$, i.e., \mbox{$\partial \chi^2 /
\partial \tilde{R}_P = 0$}. From this formula we can calculate
the estimated radius:
\begin{equation} \label{eq:R_estimate}
  \tilde{R}_P = R_\star \sqrt{- \frac{\sum_{k}\left[{\varphi_\star}_k
  \Re'_k / \left(1+\Re'_k \right) \right]}{\sum_{k} \left[{\varphi_\star}_k
  / \left( 1 + \Re'_k \right) \right]}}.
\end{equation}

Once we determine if an atmosphere is observable or not
(depending on the S/N ratio), we can use a similar approach to
quantify the detectability of the single atmospheric absorber
contributing to the total signal $\varphi$. Let
\mbox{$\hat{\varphi}_i = {\varphi_\star} (1 + \hat{\Re'}_i)$}
be the signal obtained by filtering the stellar light out of
all atmospheric absorbers except the $i^\mathrm{th}$, and let
$\tilde{\left(\hat{\varphi}_i\right)}$ be its estimation. Here,
$\hat{\Re}'_i$ is the spectrum ratio calculated when the
species $i$ is not present in the atmosphere. Further, since
$\tilde{\left(\hat{\varphi}_i\right)} \approx \alpha_i
\hat{\varphi}_i$, we can deduce the presence of absorber $i$ in
the atmosphere, by simply comparing the fit we made assuming
its absence ($\alpha_i \hat{\varphi}_i$) with the measured
signal ($\varphi$):
\begin{equation}
\chi^2_i = \sum_{k} \left( \frac{{\varphi_\star}_k}{1 + \Re'_k}
\left[ \left(1 + \Re'_k \right) - \alpha_i \left(1 +
\hat{\Re}'_{ik} \right) \right]^2 \right),
\end{equation}
where
\begin{equation}
\alpha_i = \frac{\sum_k \left[ {\varphi_\star}_k \left( 1 +
\hat{\Re}'_{ik} \right) \right]}{\sum_{k} \left[
{\varphi_\star}_k \left( 1 + \hat{\Re}'_{ik} \right)^2 / \left(
1 + \Re'_k \right) \right]}.
\end{equation}

\section{Results and discussion}
%===============================
\label{sec:results}

The results of our computations are displayed in
Tables~\ref{tab:results1} \&~\ref{tab:results2} and plotted as
spectrum ratios in Figs.~\ref{fig:ABC_ratios},
\ref{fig:DEF_ratios} \&~\ref{fig:GHI_ratios}.

\subsection{Spectral features of interest}
%-----------------------------------------
Here we summarize the contributions of each atmospheric
absorber to the spectrum ratio for various models. The spectral
resolution of the plots presented here is 10~nm. The most
prominent spectral signatures, when present, are those of O$_3$
and H$_2$O. Carbon dioxide is hard to distinguish from H$_2$O
bands and/or its own Rayleigh scattering. Molecular oxygen
transitions are too narrow to significantly contribute to the
spectrum ratio.

\subsubsection{Ozone}
%''''''''''''''''''''
In the spectral domain studied here, the Hartley (200--350~nm)
and Chappuis (420--830~nm) bands of O$_3$ appear to be the best
indicators of an Earth-like atmosphere. These bands are large
(respectively 150 and 600~nm) and lay at the blue edge of the
spectrum, where spectral features from other species are
missing. There is noticeably no contamination by H$_2$O, and
O$_2$ strong transitions are narrow and could be easily
separated. Ozone bands significantly emerge from Rayleigh
scattering and they correspond to very strong transitions,
despite the small amount of O$_3$ present in the model~A1
atmosphere (\mbox{$Y_{\mathrm{O}_3} < 10^{-5}$}). When present,
ozone is more detectable in an atmosphere similar to model~A2.

\subsubsection{Water}
%''''''''''''''''''''
The signature of H$_2$O is visible in a transit spectrum only
if H$_2$O is substantially abundant above the clouds. This is
not the case for models of Earth-like atmosphere like A1, A2
and~A3. On the contrary, the models of the ocean-planets (C1,
C2 and~C3) show a major contribution from this molecule, in the
form of four large bands that dominate the red part of the
spectrum (at $\lambda \ga 950$~nm). For these three cases,
H$_2$O can be significantly abundant above the clouds.

\subsubsection{Carbon dioxide}
%'''''''''''''''''''''''''''''
The lines of CO$_2$ are about as strongly emerging from the
`continuum' than the H$_2$O ones, but are often overlapping
with these lines. The transitions around 1\,600~nm and the ones
around 1\,950~nm are the easiest to identify, other bands are
not observable if water is present. Rayleigh scattering and
photo-absorption cross sections of CO$_2$ are comparable at
most wavelengths below 1.8~$\mu$m (see Fig.~\ref{fig:abc}),
except for a few $\sim$10-nm wide bands. In fact, the more
CO$_2$ is present in the atmosphere, the more opaque the
atmosphere becomes. This implies it would be impossible for an
observer on the surface of Venus to see the Sun. Carbon dioxide
may be more detectable farther in the infrared, hence making
desirable further investigations up to 2.5~$\mu$m.

\subsubsection{Molecular oxygen}
%'''''''''''''''''''''''''''''''
Molecular oxygen does not appear in the plots: its bands at
620, 700, 760 and 1\,260~nm are too thin to appear with only
10~nm resolution. Besides, its Rayleigh scattering cross
section almost completely masks its absorption features (see
Fig.~\ref{fig:abc}) so that no large bands of O$_2$ can be used
as an indicator of its presence. However, note that the
presence of O$_3$ indirectly indicates the presence of O$_2$,
as pointed out by L\'eger et al.\ (1993) and others.

\subsubsection{Rayleigh scattering}
%''''''''''''''''''''''''''''''''''
When Hartley and Chappuis bands of O$_3$ are absent (all cases
but A1), the Rayleigh scattering signature is clearly visible
in the blue part of the spectrum ratio. On one side it masks
the presence of some transitions, like those of O$_2$ and some
of CO$_2$, but on the other side it can provide two important
informations: (i) even if the spectral features cannot be
distinguished because they are too thin or faint, the
characteristic rising `continuum' as $\lambda^{-4}$ for short
wavelengths is a clear indication that the planet has an
atmosphere, and (ii) it indirectly indicates the presence of
the most abundant species of the atmosphere, such as CO$_2$ and
N$_2$, even if N$_2$ shows no spectral signature in the
observed domain. As a consequence, Rayleigh scattering can be
considered a way to detect N$_2$, provided clouds and/or
aerosols do not in turn mask the Rayleigh scattering signature.

To summarize, it is possible to detect the presence of the
atmosphere of a transiting exoplanet thanks to the Rayleigh
scattering, whatever the composition of the atmosphere is.
Moreover, it is theoretically possible to discriminate between
an O$_2$-rich atmosphere, where O$_3$ is expected to be present
(L\'eger et al.~1993; Sagan et al.~1993) and a H$_2$O-rich
atmosphere, as the O$_3$ lifetime is supposed to be extremely
brief in a water-rich environment. In other words, we should be
able to distinguish telluric Earth-like planets with low
volatile content from volatile-rich planets. On the other hand,
high spectral resolution is needed to discriminate between
H$_2$O-rich planets and Cytherean worlds (B1, B2, B3).

\subsection{Parameters influencing the signal-to-noise ratio}
%------------------------------------------------------------

\subsubsection{Influence of the star}
%''''''''''''''''''''''''''''''''''''
\label{sec:influence_star}

From Table~\ref{tab:results1} it is clear that the best targets
are K-type stars, rather than G- or F-type stars, the former
allowing much better S/N than the latter. Two factors are
determining the role of the star in the capabilities of
detecting an exoplanet atmosphere: (i) The size $R_\star$ of
the star, which directly influences the S/N (see
Eq.~\ref{eq:chi2}) and the duration of transit
(Eq.~\ref{eq:time_transit}) and (ii) the semi-major axis of the
planet's orbit, which influences both the duration of transit
and the probability to observe the transit from Earth (see
below). These factors can explain the discrepancies between the
S/N values obtained for different kind of stars in
Table~\ref{tab:results1}.

The probability, $\alpha$, that a planet transiting its parent
star might be seen from the Earth is defined as $\alpha \equiv
P\{{\rm transit}\} = R_\star / a $, with $R_\star$ being the
radius of the star and $a$ the semi-major axis of the planet's
orbit. This probability is about 10\% for `hot Jupiters', while
it is 0.3\%, 0.5\% and 0.7\% for planets orbiting in the HZ of
a F, G or K star, respectively.

In addition, K stars are more numerous than other types of
stars. From the CDS database, we find there is approximatively
a total of $10\,000 \cdot 10^{0.6(V-8)}$ main sequence stars
brighter than a given magnitude $V$ on the whole
sky.\footnote{We consider mostly bright stars, for which the
distribution is essentially isotropic.} About $3/5$ of these
are K type stars, against only $1/10$ for G stars. Let us now
define $\beta$ to be the number of planet(s) per star, and
$\gamma$ to be the fraction of the sky that is considered for a
transit detection survey (in other words the efficiency of
surveys to find the targets). We list in
Table~\ref{tab:results2} the number of potential targets for
each model. This number, $N$, corresponds to the number of
targets detected with a 10-m telescope mirror effective size
and with a S/N greater than or equal to 5. It is given by:
\begin{equation}
\label{eq:N_computed} N_{\mathrm{S/N} \geq 5,~\epsilon D =
10\mathrm{\,m}} = N_0 \cdot \alpha \cdot \beta \cdot \gamma
\cdot \left(\frac{\mathrm{S/N}_{V=8,\,\epsilon
D=10\mathrm{\,m}}}{5}\right)^3,
\end{equation}
where $N_0$ is about 6\,000, 1\,000 and 3\,000 for K, G and F
stars respectively, i.e. the number of stars with magnitude
$\geq 8$, and S/N$_{V=8,\,\epsilon D=10\,m}$ is the expected
S/N ratio computed for a given atmosphere of a planet orbiting
a $V=8$ star with a telescope having a mirror effective area of
10~m (this value is given in the last column of
Table~\ref{tab:results1}). Since no Earth-size planet has been
discovered so far, we have no real estimate of $\beta$. In the
following, when it is not a free parameter we consider
$\beta=1$\footnote{Actually, $\beta = 2$ in the Solar System
because there are two Earth-size planets with atmospheres,
namely Venus and the Earth.}. Catala et al.\ (2005) propose a
$30\degr \times 30\degr$ survey dedicated to find planets
around $< 11^\mathrm{th}$-magnitude stars, i.e., $\gamma
\approx$ 2--3\% for such a project.

Let be $N_{\mathrm{S/N},~\epsilon D}$, the number of potential
targets reaching a minimum S/N ratio for a given mirror
effective size $\epsilon D$, which scales from the value
calculated using Eq.~\ref{eq:N_computed}, $N_{\mathrm{S/N} \geq
5, \epsilon D = 10\mathrm{\,m}}$, in the following way:
\begin{equation}
\label{eq:N_scaled} N_{\mathrm{S/N},~\epsilon D} =
N_{\mathrm{S/N}\geq 5,~\epsilon D=10\mathrm{\,m}} \cdot
\left(\frac{\mathrm{S/N}}{5}\right)^{-3} \cdot
\left(\frac{\epsilon D}{10\mathrm{~m}}\right)^3.
\end{equation}

The values obtained for atmospheric detection are strongly in
favor of a small, late type star. Note that this is also true
for the detection of the planetary transit as well.

\subsubsection{Effect of the atmospheric temperature gradient}
%'''''''''''''''''''''''''''''''''''''''''''''''''''''''''''''
\label{sec:temperature_effect}

The thick CO$_2$ Venus-like atmospheres (B1, B2 and~B3, see
Table~\ref{tab:results1} \&~\ref{tab:results2}) are more
difficult to detect than other cases. Even if we set the top of
the clouds at 10~km height, the detection remains more
challenging than for model~A1. That is somewhat surprising,
partly because CO$_2$ has strong transitions, particularly in
the near infrared, and partly because of the larger scale
height at the surface of the planet (14.3~km for model~B1,
8.8~km for model~A1). As a consequence, the atmosphere in
model~B1 should have a larger vertical extent than in model~A1.
In reality, the difficulty to characterize the atmospheres of
models~B1, B2 and~B3 is related to the temperature profiles we
chose (see Fig.~\ref{fig:A_profile} and~\ref{fig:B_profile}):
At 50~km of altitude, the temperature of model~B1 is roughly
60~K colder than that of model~A1. This model in fact benefits
from the positive stratospheric temperature gradient of the
Earth. Moreover, the atmosphere for model~B1
($\mu_\mathrm{B1}=43$~g\,mol$^{-1}$) is heavier than the one
for model~A1 ($\mu_\mathrm{A1}=29$~g\,mol$^{-1}$). Therefore,
at high altitude, the scale height is larger in model~A1 than
in model~B1 (respectively 7.6~km and 3.9~km at an altitude of
50~km).

\subsubsection{Effect of atmospheric pressure}
%'''''''''''''''''''''''''''''''''''''''''''''
\label{sec:pressure_effect}

Note that the thickness of the atmosphere in model~B1 is almost
half the one in A1, despite the intense surface pressure
(100~atm), which should help to increase the upper level of the
atmosphere, limited by the UV photo-dissociation
($h_\mathrm{max}$). The exponential decrease of pressure
prevents, in fact, $p_0$ to play a key role: in order to
counterbalance the effect of the negative temperature gradient,
the surface pressure should have been $>10^6$~atm to obtain
absorptions similar to the case of the Earth (model~A1).

\subsubsection{Effect of the planet gravity and density}
%'''''''''''''''''''''''''''''''''''''''''''''''''''''''
The atmospheric absorption is, at a first order, proportional
to $H \cdot R_P$. At a given temperature and for a given
atmospheric composition, the scale height $H$ is proportional
to the inverse of the gravity acceleration, $g^{-1}$, or
equivalently to $R_P^2/M_P$, where $M_P$ is the mass of the
planet. As a result, the absorption is expected to be roughly
inversely proportional to the bulk density of the planet,
$\rho_P$, independently of the planet size.

\begin{figure}[htbp!]
    \resizebox{\hsize}{!}{\includegraphics{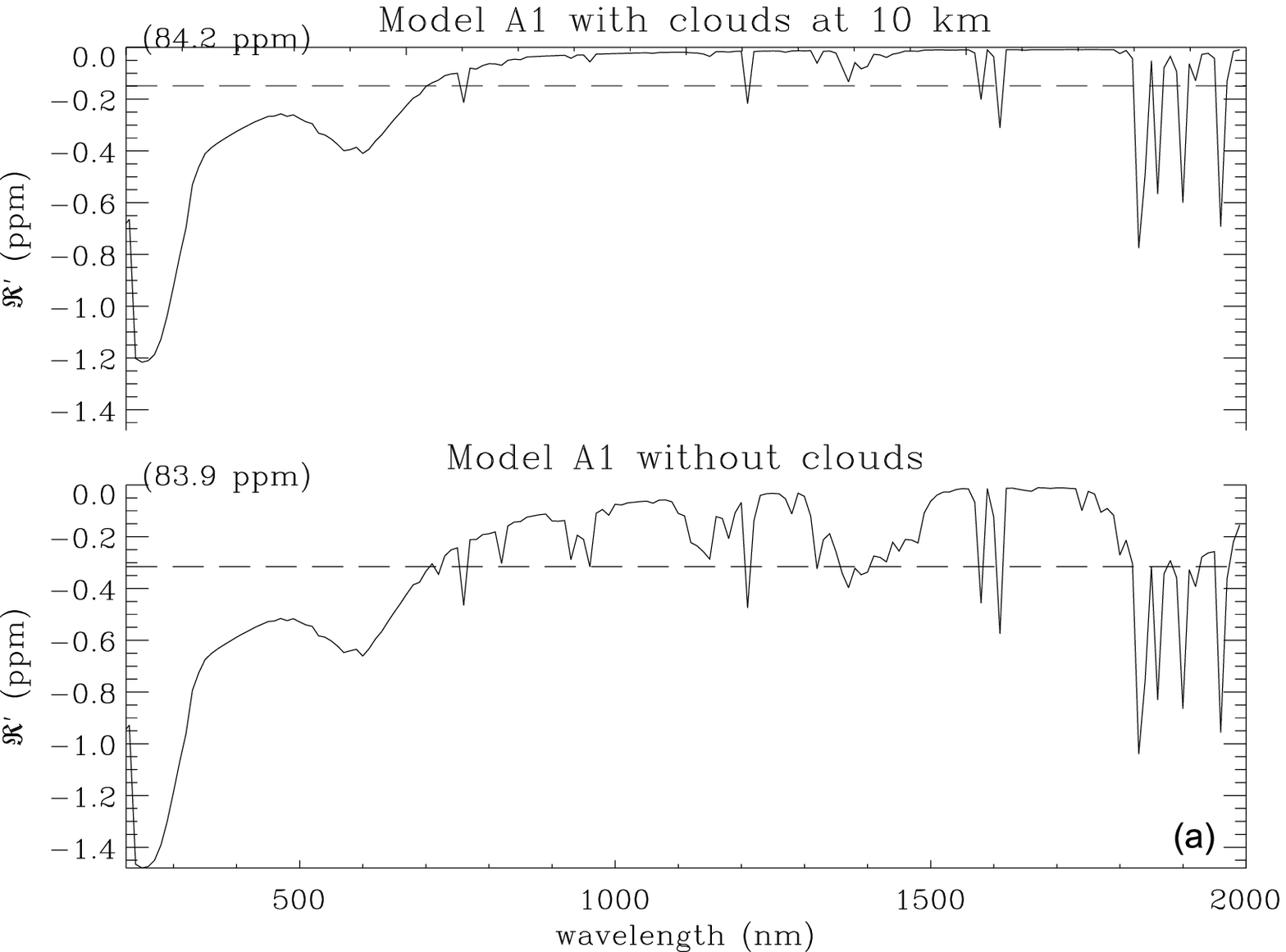}}
    \resizebox{\hsize}{!}{\includegraphics{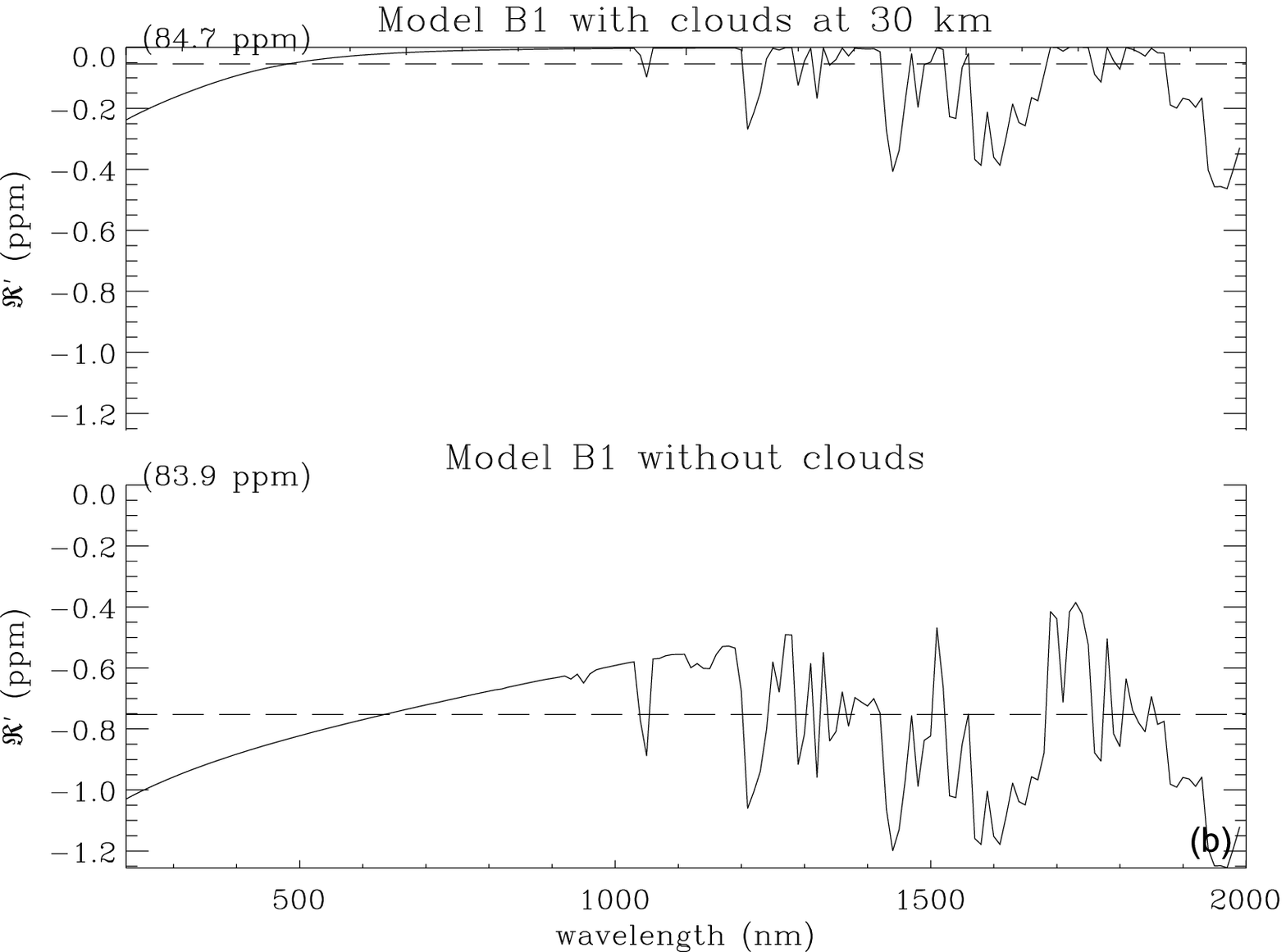}}
    \resizebox{\hsize}{!}{\includegraphics{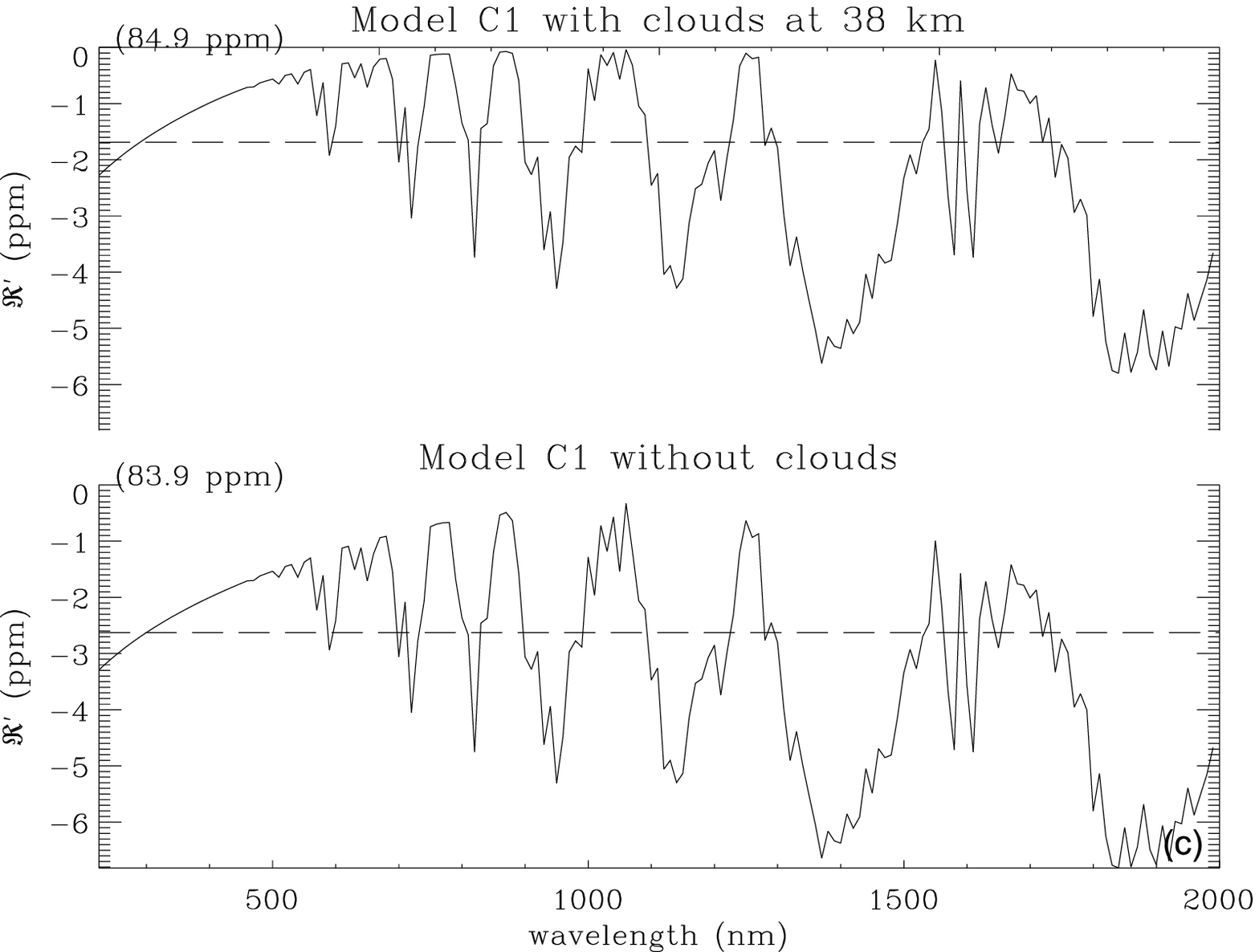}}
    \caption{Spectrum ratios for models~A1 (a), B1 (b) and~C1 (c). The
    spectrum ratios have been respectively shifted by the values in
    parenthesis so that the absorption by the `solid disk' of
    the planet is 0~ppm. In the case of models with clouds, the
    `solid disk' is artificially increased by the cloud layer. The
    dashed line indicates the best fit estimation of the radius
    of the planet, $\tilde{R}_P$ (see Sect.~\ref{sec:S/N}) if we
    suppose there is no atmosphere.}
    \label{fig:ABC_ratios}
\end{figure}

\begin{figure}[htbp!]
    \resizebox{\hsize}{!}{\includegraphics{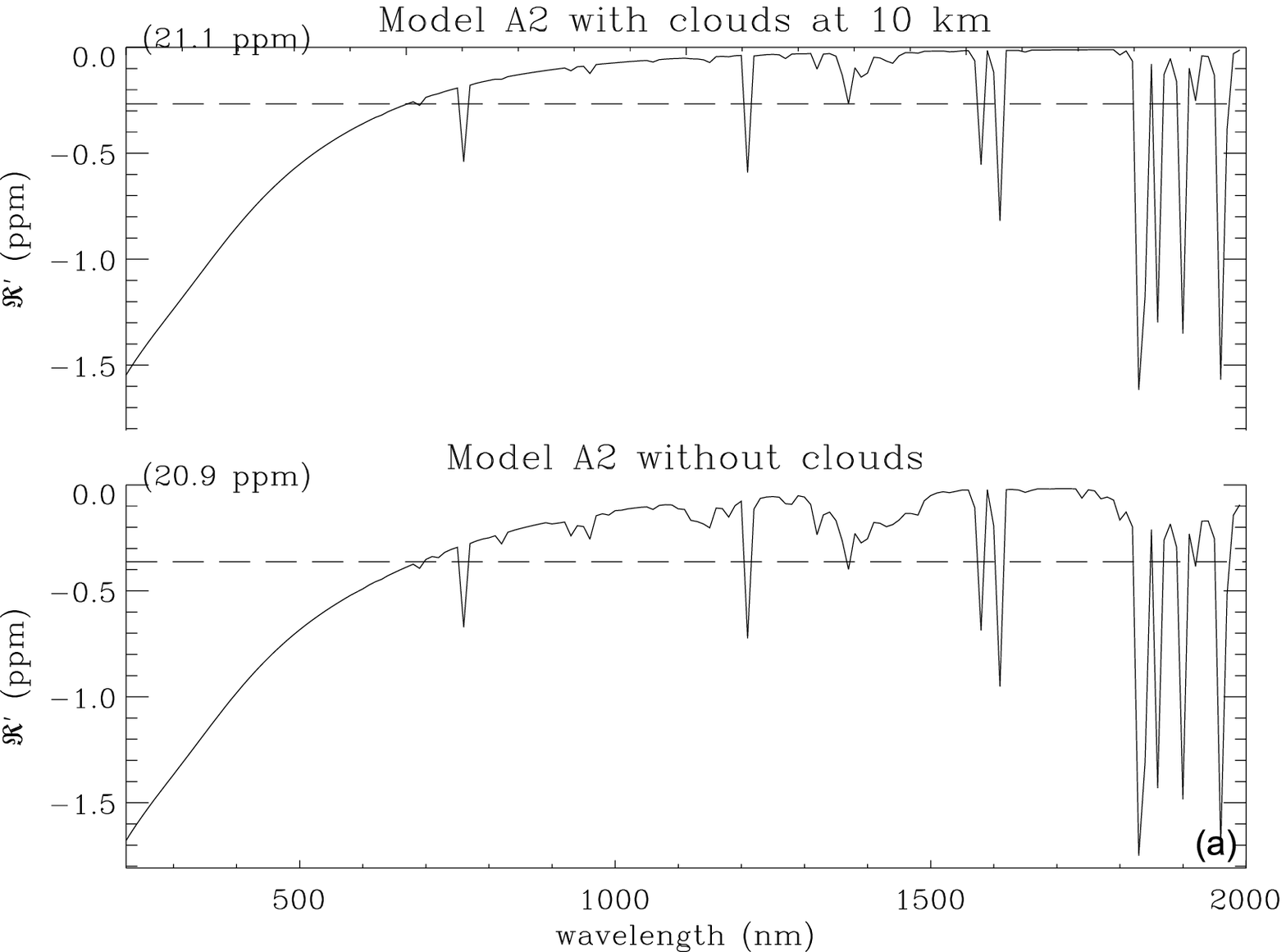}}
    \resizebox{\hsize}{!}{\includegraphics{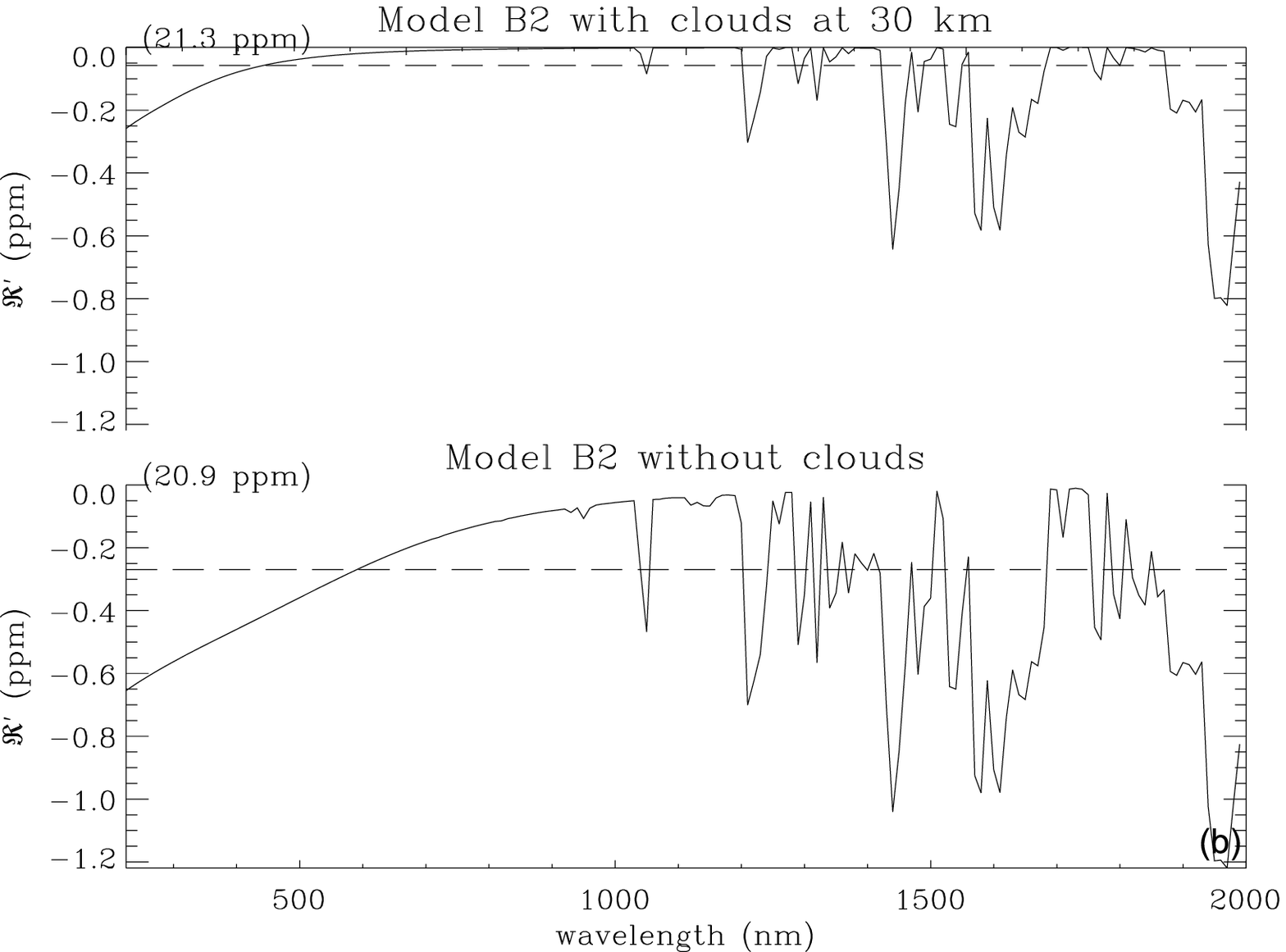}}
    \resizebox{\hsize}{!}{\includegraphics{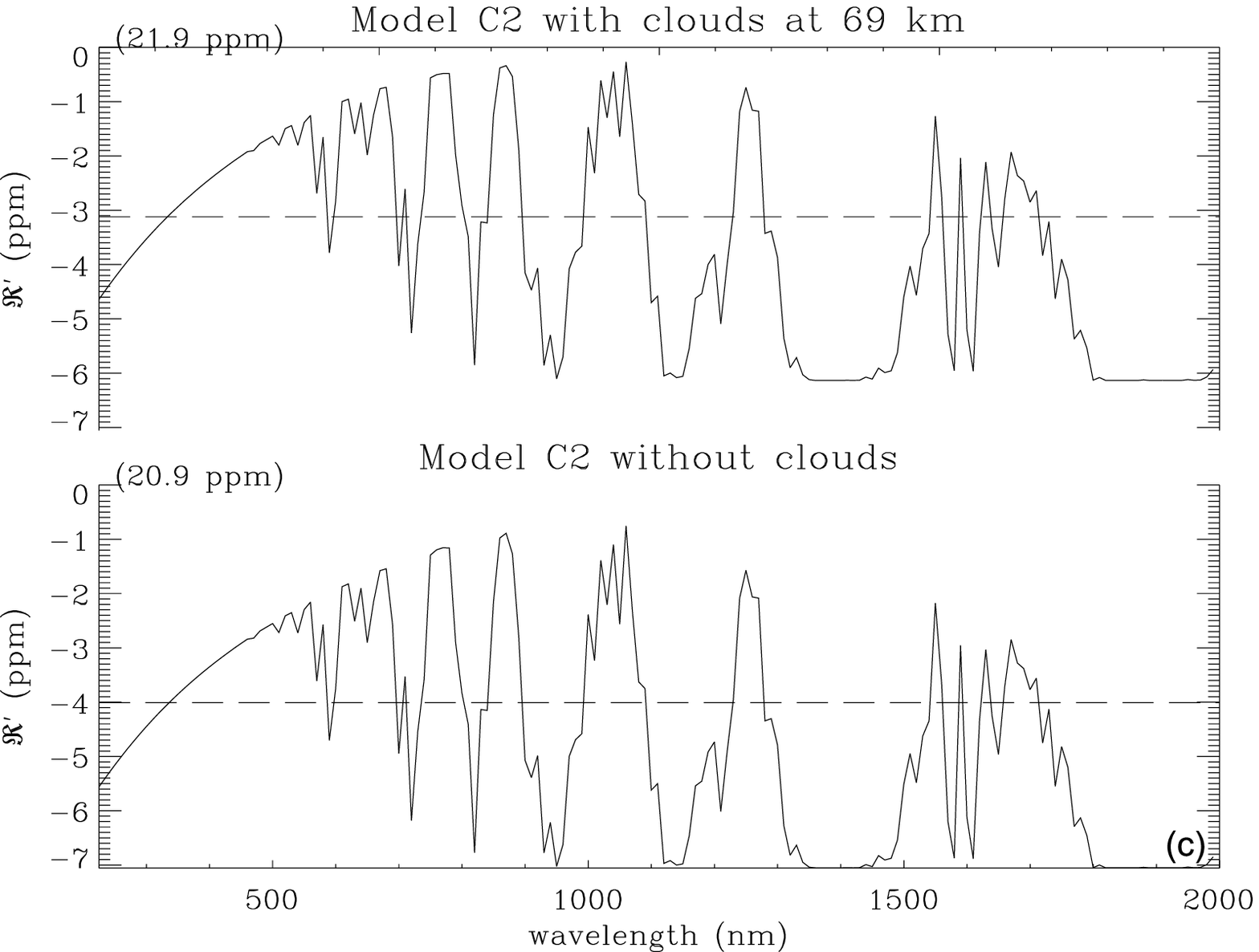}}
    \caption{Spectrum ratios for models~A2 (a), B2 (b) and~C2 (c).
    The `saturation effect' in H$_2$O lines, for model~C2, is a consequence of the
    atmosphere being optically thick at the upper atmospheric level,
    $h_\mathrm{max}$. In fact, if one consider there is no more
    water above this level due to photo-dissociation (see
    Sect.~\ref{sec:b_max}), such transmitted spectrum plots
    allow to determine the level where H$_2$O photo-dissociation
    occurs in an exoplanet atmosphere.}
    \label{fig:DEF_ratios}
\end{figure}

\begin{figure}[htbp!]
    \resizebox{\hsize}{!}{\includegraphics{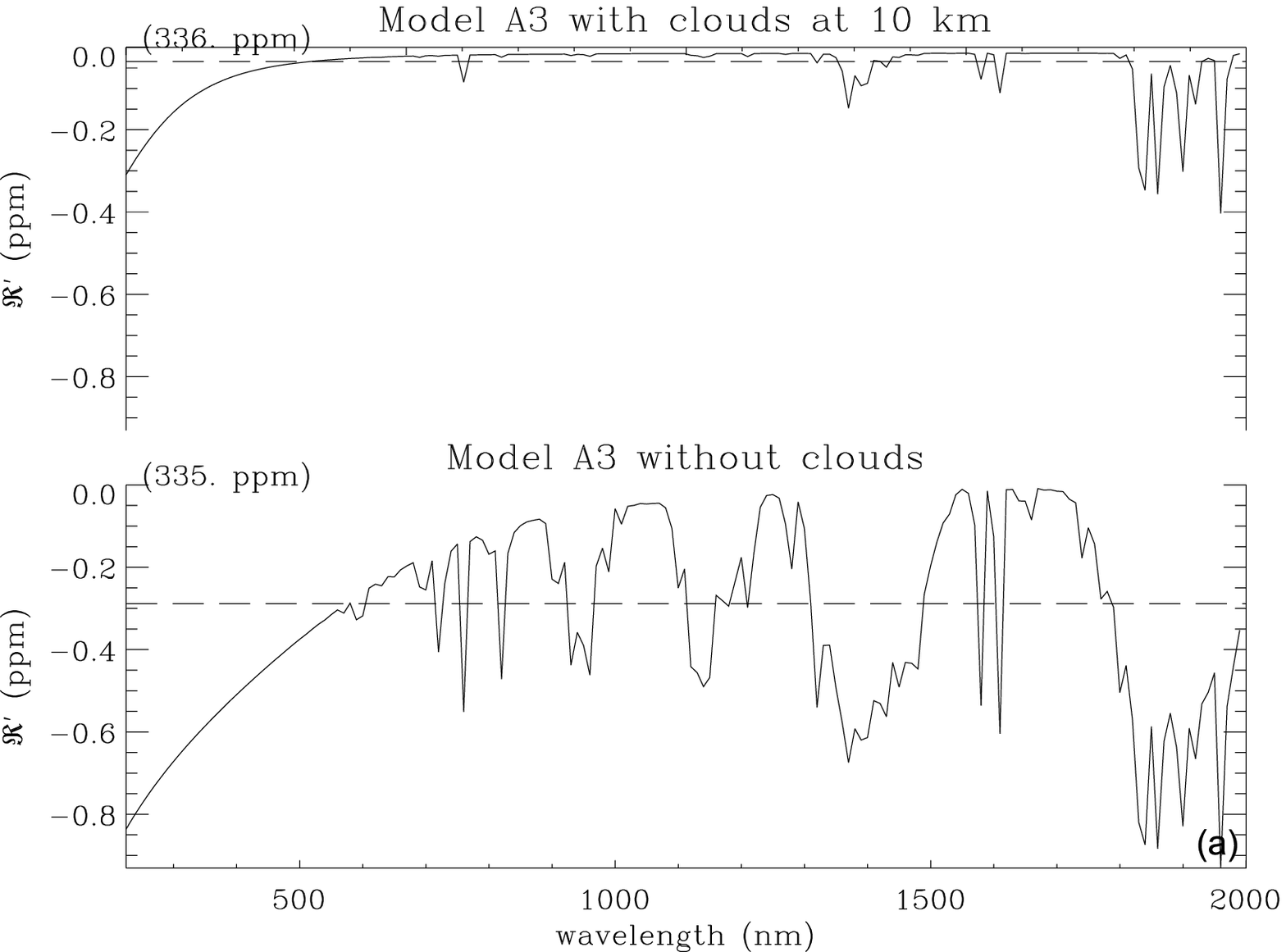}}
    \resizebox{\hsize}{!}{\includegraphics{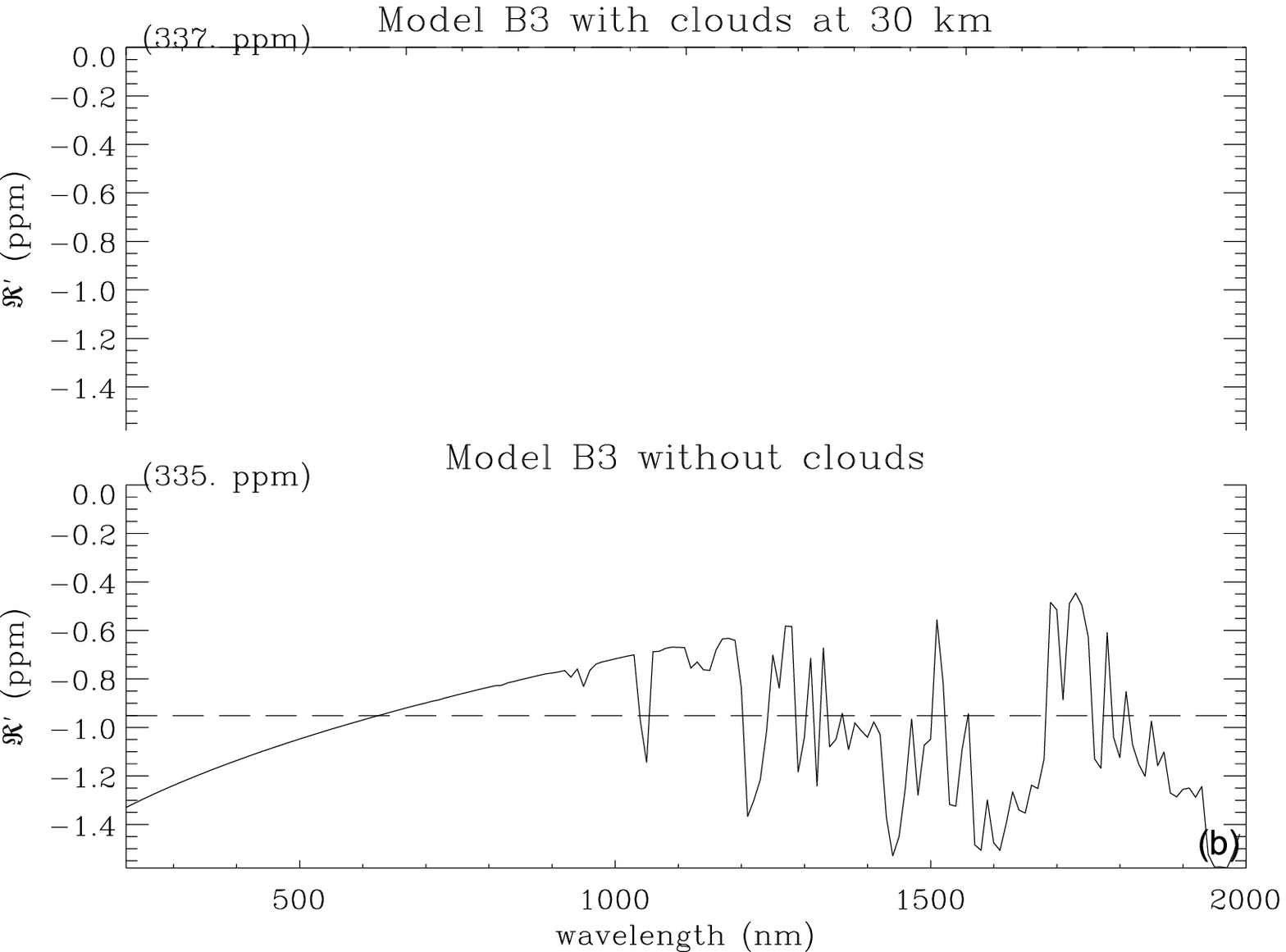}}
    \resizebox{\hsize}{!}{\includegraphics{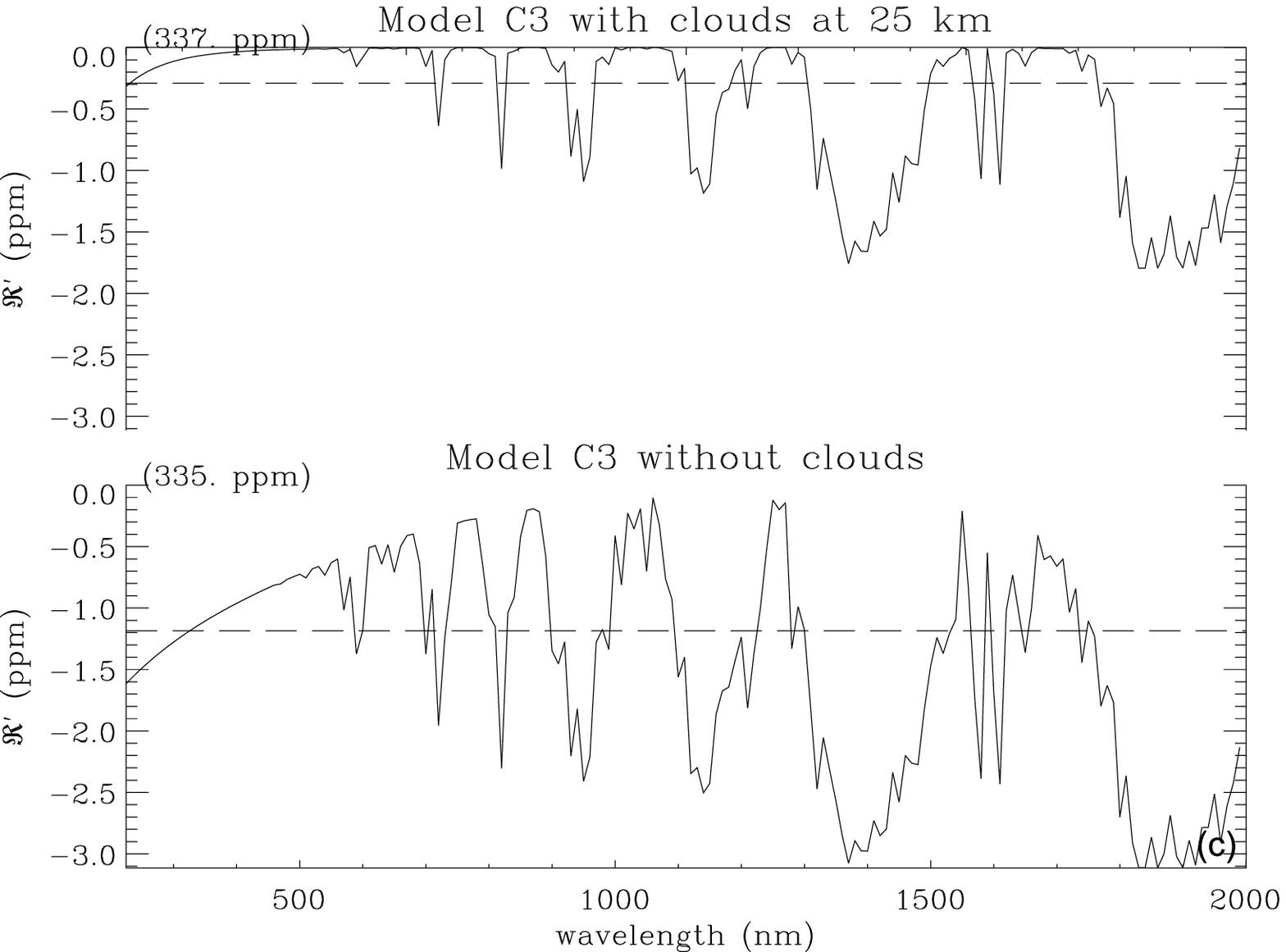}}
    \caption{Spectrum ratios for models~A3 (a), B3 (b) and~C3 (c).}
    \label{fig:GHI_ratios}
\end{figure}

This effect is illustrated by the following examples:
models~C1, C2 and~C3 all benefit from very extended
atmospheres, given the weak value of $g$ in the three cases.
For a planet as dense as the Earth (such that $g_\mathrm{C1} =
g_\mathrm{A1}$), the results for the N$_2$/H$_2$O-rich
atmosphere in models~C are close to the ones obtained for
models~A. Both models~C and~A, present typical spectral
features. In model~A1, ozone, for which the concentration peaks
at the tropopause, gives a prominent signature in the blue edge
of the spectral domain (the Hartley and Chappuis bands, as seen
in Fig.~\ref{fig:ABC_ratios}, top panel). On the contrary, the
saturated atmosphere of model~C1, which sustain H$_2$O up to
high altitudes, yields strong bands around 0.14 and 0.19~$\mu$m
(Fig.~\ref{fig:ABC_ratios}, bottom panel). The role played by
$g$ can be better understood by comparing model~A3 or~B3
($g=24.5$~m\,s$^{-2}$) to model~A2 or~B2 ($g=3.9$~m\,s$^{-2}$),
and model~C3 ($g=14.7$~m\,s$^{-2}$) to model~C2
($g=2$~m\,s$^{-2}$). Using absorption spectroscopy, it is clear
that the atmospheres of small and light planets (i.e., with low
surface gravity) are easier to detect than the ones of large
and dense planets (i.e., with high surface gravity).

Small and light exoplanets, however, may not be able to retain
a thick atmosphere. In fact, high thermal agitation of
atmospheric atoms causes particles to have a velocity in the
tail of the Maxwellian distribution allowing them to escape
into space (i.e., Jean's escape). It is therefore questionable
if planets of the size of Titan can have a dense atmosphere at
1~AU from their star. Models~A2, B2 and~C2 enter that category.
This problem concerns both small planets and giant exoplanet
satellites.

According to Williams et al.\ (1997), a planet having the
density of Mars could retain N and O over more than 4.5~Gyr if
it has a mass greater than 0.07~M$_\oplus$. Model planets~A2
and B2 have masses of 0.1~M$_\oplus$ and a density equivalent
to that of Mars ($\approx$4~g\,cm$^{-3}$) so they would be able
to retain an atmosphere (though they may not be able to have a
1~atm atmosphere, as for Mars). The ocean-planet model~C2 has a
mass of 0.05~M$_\oplus$ for a density  of 2.8~g\,cm$^{-3}$, and
according to Williams et al.\ (1997), its atmosphere should
consequently escape. However, although at 1~AU from the star,
such a planet also has a huge reservoir of volatile elements.
This reservoir should help to `refill' the escaping atmosphere.

Note that an hydrodynamically escaping atmosphere should be
easier to detect than a stable one, since it can bring heavier
elements into the hot upper atmosphere. This effect is
illustrated by the absorptions seen by Vidal-Madjar et al.\
(2003, 2004) in the spectrum of \object{HD~209\,458}, which
originate in its transiting giant planet hydrodynamically
escaping atmosphere. A model of an `escaping ocean' is studied
by Jura (2004). This process would give interesting absorption
signatures in the H$_2$O bands from the lower atmosphere and in
the signatures of the photo-dissociation products of H$_2$O
from the upper atmosphere (such as an absorption of
Lyman~$\alpha$ photons by the hydrogen atom). See detailed
discussion in Jura (2004).

\begin{table*}
    \centering
    \begin{tabular}{*{10}{c}}
        \hline \hline
        Model & Description         & Atm. type          & $R_P$        & $M_P$        & $\rho_P$       & $g$           & $p_0$ & $H_0$ & $h_\mathrm{max}$ \\
              &                     &                    & (R$_\oplus$) & (M$_\oplus$) & (g\,cm$^{-3}$) & (m\,s$^{-2}$) & (atm) & (km)  & (km)\\
        \hline
        A1     & ($\approx$)Earth    & N$_2$/O$_2$-rich   & 1            & 1            & 5.5            & 9.8           & 1     & 8.8   & 85  \\
        B1     & ($\approx$)Venus    & CO$_2$-rich        & 1            & 1            & 5.5            & 9.8           & 100   & 14.3  & 50  \\
        C1     & medium ocean-planet & N$_2$/H$_2$O-rich  & 1            & 0.5          & 2.8            & 4.9           & 1     & 20.0  & 260 \\
        A2     & small Earth         & N$_2$/O$_2$-rich   & 0.5          & 0.1          & 4.0            & 3.9           & 1     & 24.7  & 260 \\
        B2     & small Venus         & CO$_2$-rich        & 0.5          & 0.1          & 4.0            & 3.9           & 1     & 40.0  & 99  \\
        C2     & small ocean-planet  & N$_2$/H$_2$O-rich  & 0.5          & 0.05         & 1.8            & 2.0           & 1     & 61.4  & 499 \\
        A3     & `super-Earth'       & N$_2$/O$_2$-rich   & 2            & 9            & 6.1            & 24.5          & 1     & 3.9   & 30  \\
        B3     & `super-Venus'       & CO$_2$-rich        & 2            & 6            & 6.1            & 24.5          & 100   & 6.4   & 30  \\
        C3     & big ocean-planet    & N$_2$/H$_2$O-rich  & 2            & 9            & 4.1            & 14.7          & 1     & 6.7   & 60  \\
        \hline
    \end{tabular}
    \caption{Summary of test models.}
    \label{tab:models}
\end{table*}

\begin{table*}
    \centering
    \begin{tabular}{*{9}{c}}
        \hline \hline
        Model   & Description      & Star & \multicolumn{6}{c}{Signal-to-noise ratio}                \\
                &                  &      & \multicolumn{6}{c}{(S/N)$_{V=8,\,\epsilon D=10\rm{~m}}$} \\
                &                  &      & w/o cloud & w/ clouds & H$_2$O & CO$_2$ & O$_3$ & O$_2$  \\
        \hline
                &                  & K    & 5.2       & 3.5       & 1.7    & 1.1    & 1.9   & 0.2    \\
        A1      & ($\approx$)Earth & G    & 3.2       & 2.3       & 0.8    & 0.5    & 1.2   & 0.2    \\
                &                  & F    & 2.3       & 1.7       & 0.5    & 0.3    & 0.9   & 0.1    \\
        \hline
                &                  & K    & 4.0       & 2.3       & 0.0    & 2.3    & -     & -      \\
        B1      & ($\approx$)Venus & G    & 2.1       & 1.2       & 0.0    & 1.2    & -     & -      \\
                &                  & F    & 1.3       & 0.7       & 0.0    & 0.7    & -     & -      \\
        \hline
                & medium           & K    & 41        & 39        & 39     & 11     & -     & -      \\
        C1      & ocean-           & G    & 22        & 20        & 20     & 5.4    & -     & -      \\
                & planet           & F    & 14        & 13        & 13     & 3.3    & -     & -      \\
        \hline
                &                  & K    & 6.9       & 6.3       & 3.8    & 2.8    & -     & 0.7    \\
        A2      & small Earth      & G    & 4.3       & 4.0       & 1.8    & 1.4    & -     & 0.5    \\
                &                  & F    & 3.2       & 3.0       & 1.1    & 0.8    & -     & 0.3    \\
        \hline
                &                  & K    & 5.8       & 3.3       & 0.0    & 3.3    & -     & -      \\
        B2      & small Venus      & G    & 3.0       & 1.6       & 0.0    & 1.7    & -     & -      \\
                &                  & F    & 1.9       & 1.0       & 0.0    & 1.0    & -     & -      \\
        \hline
                & small            & K    & 47        & 46        & 46     & 17     & -     & -      \\
        C2      & ocean-           & G    & 26        & 25        & 25     & 8.6    & -     & -      \\
                & planet           & F    & 17        & 16        & 16     & 5.2    & -     & -      \\
        \hline
                &                  & K    & 4.6       & 1.1       & 0.9    & 0.5    & -     & 0.1    \\
        A3      & super-Earth      & G    & 2.5       & 0.6       & 0.4    & 0.2    & -     & 0.1    \\
                &                  & F    & 1.7       & 0.4       & 0.3    & 0.1    & -     & 0.0    \\
        \hline
                &                  & K    & 5.6       & 0         & 0      & 0      & -     & -      \\
        B3      & super-Venus      & G    & 2.9       & 0         & 0      & 0      & -     & -      \\
                &                  & F    & 1.9       & 0         & 0      & 0      & -     & -      \\
        \hline
                & big              & K    & 20        & 13        & 12     & 3.2    & -     & -     \\
        C3      & ocean-           & G    & 10        & 6.5       & 6.3    & 1.5    & -     & -     \\
                & planet           & F    & 6.7       & 4.1       & 4.0    & 0.9    & -     & -     \\
        \hline
    \end{tabular}
    \caption{Summary of results: signal-to-noise ratios obtainable with
    a telescope mirror effective size of $\epsilon D = 10$~m pointing at a $V=8$ star.
    To get the S/N ratios for a different effective size $\epsilon D$, exposure time during transit, $t$,
    and/or apparent magnitude of the star, $V$, the result
    scales with \mbox{$(\epsilon D / 10\mathrm{~m}) \cdot (t/\tau)^{0.5} \cdot 10^{-0.2 (V - 8)}$}
    where $\tau$ is defined by Eq.~\ref{eq:time_transit}. The S/N by species are calculated for the models with clouds.}
    \label{tab:results1}
\end{table*}

\begin{table*}
    \centering
     \begin{tabular}{*{9}{c}}
        \hline \hline
        Model   & Description      & Star & Mirror        & Limiting  & Number   & \multicolumn{3}{c}{Number~of~targets}                                               \\
                &                  &      & eff. size (m) & magnitude & of stars & \multicolumn{3}{c}{for models w/ clouds}                                                                                    \\
                &                  &      & $(\epsilon D)_{\mathrm{S/N}\geq5,\,V=8}$ & $(V_\mathrm{Lim})_{\mathrm{S/N}\geq5,\,\epsilon D=10\mathrm{\,m}}$ &          & \multicolumn{3}{c}{$(N)_{\mathrm{S/N}\geq5}$, $\epsilon=50\%$} \\
                &                  &      & w/ clouds     & w/ clouds &          & $\beta \cdot \gamma = 1$ & $\beta \cdot \gamma = 3\%$ & $\beta \cdot \gamma = 10\%$ \\
                &                  &      &               &           &          & $D = 20$~m               & D = $30$~m                 & $D = 30$~m                  \\
        \hline
                &                  &      & (a)      & (b)  & (c)           & \multicolumn{3}{c}{(d)} \\
        \hline
%       Model   & Description      & Star & Diameter & Mag. & N stars       & Ntarget(1) & Ntarget(2) & Ntarget(3) \\
                &                  & K    & 14       & 7.22 & 2\,042        & 14         & 1          & 4          \\
        A1      & ($\approx$)Earth & G    & 22       & 6.31 & 96            & $<1$ (0.4) & $\ll 1$    & $<1$ (0.1) \\
                &                  & F    & 29       & 5.66 & 118           & $<1$ (0.3) & $\ll 1$    & $<1$ (0.1) \\
        \hline
                &                  & K    & 21       & 6.31 & 580           & 4          & $<1$ (0.4) & 1          \\
        B1      & ($\approx$)Venus & G    & 43       & 4.90 & 13            & $\ll 1$    & $\ll 1$    & $\ll 1$    \\
                &                  & F    & 68       & 3.73 & 8             & $\ll 1$    & $\ll 1$    & $\ll 1$    \\
        \hline
                & medium           & K    & 1.3      & 12.5 & $>3\cdot10^6$ & 19\,602    & 1\,984     & 6\,615     \\
        C1      & ocean-           & G    & 2.5      & 11.0 & 63\,095       & 321        & 32         & 108        \\
                & planet           & F    & 3.9      & 10.1 & 54\,591       & 157        & 15         & 52         \\
        \hline
                &                  & K    & 8        & 8.50 & 11\,971       & 84         & 8          & 28         \\
        A2      & small Earth      & G    & 13       & 7.51 & 508           & 2          & $<1$ (0.2) & $<1$ (0.6) \\
                &                  & F    & 17       & 6.90 & 656           & 1          & $<1$ (0.1) & $<1$ (0.3) \\
        \hline
                &                  & K    & 15       & 7.10 & 1\,730        & 12         & 1          & 4          \\
        B2      & small Venus      & G    & 31       & 5.52 & 32            & $<1 (0.1)$ & $\ll 1$    & $\ll 1$    \\
                &                  & F    & 50       & 4.50 & 23            & $\ll 1$    & $\ll 1$    & $\ll 1$    \\
        \hline
                & small            & K    & 1.1      & 12.8 & $>4\cdot10^6$ & 33\,569    & 3\,398     & 11\,329    \\
        C2      & ocean-           & G    & 2.0      & 11.5 & 125\,892      & 600        & 60         & 202        \\
                & planet           & F    & 3.1      & 10.5 & 94\,868       & 307        & 31         & 103        \\
        \hline
                &                  & K    & 45       & 4.71 & 63            & $<1$ (0.4) & $\ll 1$    & $<1$ (0.1) \\
        A3      & super-Earth      & G    & 86       & 3.39 & 1             & $\ll 1$    & $\ll 1$    & $\ll 1$    \\
                &                  & F    & 121      & 2.51 & 1             & $\ll 1$    & $\ll 1$    & $\ll 1$    \\
        \hline
                &                  & K    & $> 10^3$ & -    & 0             & $\ll 1$    & $\ll 1$    & $\ll 1$    \\
        B3      & super-Venus      & G    & $> 10^3$ & -    & 0             & $\ll 1$    & $\ll 1$    & $\ll 1$    \\
                &                  & F    & $> 10^3$ & -    & 0             & $\ll 1$    & $\ll 1$    & $\ll 1$    \\
        \hline
                & big              & K    & 4.0      & 10.1 & 109\,182      & 682        & 69         & 230        \\
        C3      & ocean-           & G    & 7.7      & 8.57 & 2\,197        & 10         & 1          & 3          \\
                & planet           & F    & 13       & 7.57 & 1\,656        & 4          & $<1$ (0.4) & 1          \\
        \hline
    \end{tabular}
    \caption{Summary of results: mirror effective size and number of targets. \newline
        (a) Effective size $(\epsilon D)_{\mathrm{S/N}\geq5,\,V=8}$ of the telescope mirror required to
        obtain $\mathrm{S/N}=5$ for a $V=8$ star, based on the
        numbers displayed for the models with clouds (see
        Table~\ref{tab:results1}).\newline
        (b) The limiting magnitude at which the number
        of targets in the last column is given. This can be expressed as
        \mbox{$(V_\mathrm{Lim})_{\mathrm{S/N}\geq5,\,\epsilon D=10\mathrm{\,m}} = 5 \cdot \log_{10} \left[ \left(\mathrm{S/N}_{V=8,\,\epsilon D = 10}\right) / 5 \cdot (\epsilon D) / 10\mathrm{~m}
        \right]+8$}. \newline
        (c) Total number of given spectral-type stars brighter than the limiting
        magnitude.\newline
        (d) Number of potential targets calculated with
        Eq.~\ref{eq:N_computed}, using the S/N value of the
        models with clouds and assuming various $\beta \cdot \gamma$
        values.
        The coefficients $\beta$ and $\gamma$ are defined
        in the text. When the number of potential targets is
        slightly less than 1, the value is given between
        parenthesis. Use Eq.~\ref{eq:N_scaled} to scale the
        value displayed in the column to any mirror
        effective size $\epsilon D$ and minimum S/N.}
    \label{tab:results2}
\end{table*}

\section{Conclusion}
%===================

The vertical extent of the atmosphere is of extreme importance
as concerns the detectability of a remote atmosphere by
absorption spectroscopy. This tends to favor less dense
objects, like giant exoplanet satellites (as would be an
`exo-Titan') or volatile-rich planets (as ocean-planets,
theoretically possible but not observed yet). Cytherean
atmospheres are the most challenging to detect. Surface
parameters, such as surface pressure and temperature, are not
crucial. A temperature gradient that becomes positive at few
tens of kilometers height (for instance owing to
photochemistry) might help the detection. Our results show that
late-type stars are better for detecting and characterizing the
atmospheres of planets in transit, since they are smaller, more
numerous and present a better probability of being transited by
a planet.

The strongest signatures of the atmosphere of a transiting
Earth-size planet could be those of H$_2$O (6~ppm in the case
of hypothetical ocean-planets), O$_3$ ($\sim$1--2~ppm) and
CO$_2$ (1~ppm), considering our spectral study from the UV to
the NIR (i.e., from 0.2 to 2~$\mu$m). The presence of an
atmosphere around hundreds of hypothetical `ocean-planets'
(models C) could be detected with a 10--20~m telescope. The
atmospheres of tens of giant exoplanet satellites (model A2)
could be in the range of a 20--30~m instrument. A 30--40~m
telescope would be required to probe Earth-like atmospheres
around Earth-like planets (model A1). These numbers suppose
that Earth-size planets are frequent and are efficiently
detected by surveys.

Finally, planets with an extended upper atmosphere, like the
ones described by Jura (2004), hosting an `evaporating ocean',
or the planets in an `hydrodynamical blow-off state', are the
natural link between the planets we have modelled here and the
observed `hot Jupiters'.

\begin{acknowledgement}
We warmly thank Chris Parkinson for careful reading and
comments that noticeably improved the manuscript, David Crisp
for the code \texttt{LBLABC} and the anonymous referee for
thorough reading and useful comments on the manuscript. G.
Tinetti is supported by NASA Astrobiology Institute -- National
Research Council.
\end{acknowledgement}

% Biblio
% ======

\end{document}